\documentclass[prb,twocolumn,superscriptaddress,showpacs,amsmath,amssymb]{revtex4-2}

\usepackage{graphicx}
\usepackage{latexsym}
\usepackage{amsmath}
\usepackage{amssymb}
\usepackage{amsfonts}
\usepackage{changes}
\usepackage{color}
\usepackage{bm}
\usepackage{verbatim}
\usepackage{pagecolor,lipsum}
\usepackage{lineno}
\usepackage[version=3]{mhchem} 
\usepackage{xcolor,cancel}

\usepackage{framed}
\definecolor{shadecolor}{RGB}{222,222,221}
\begin{document}

\title{Magnon influence on the superconducting density of states in
    superconductor--ferromagnetic-insulator bilayers
}

\author{A.S. Ianovskaia}
\affiliation{Moscow Institute of Physics and Technology, Dolgoprudny, 141700 Russia}
\affiliation{National Research University Higher School of Economics, Moscow, 101000 Russia}

\author{A.M. Bobkov}
\affiliation{Moscow Institute of Physics and Technology, Dolgoprudny, 141700 Russia}

\author{I.V. Bobkova}
\affiliation{Moscow Institute of Physics and Technology, Dolgoprudny, 141700 Russia}
\affiliation{National Research University Higher School of Economics, Moscow, 101000 Russia}

\date{\today}


\begin{abstract}
Superconductor--ferromagnetic-insulator heterostructures are paradigmatic systems for studying the mutual influence of superconductivity and magnetism via proximity effects. In particular, spin-split superconductivity is realized in such structures. Recent experiments and theories demonstrate a rich variety of transport phenomena occurring in devices based on such heterostructures that suggest direct applications in thermoelectricity, low-dissipative spintronics, radiation detection, and sensing. In this work we investigate the influence of the electron-magnon interaction at the superconductor--ferromagnetic-insulator interface on the spin-split  superconductivity. It is predicted that due to the magnon-mediated electron spin-flip processes the spin-split quasiparticle branches are partially mixed and reconstructed, and the BCS-like spin-split shape of the superconducting density of states, which is typical for superconductors in the effective exchange field,  is strongly modified. An odd-frequency superconducting order parameter admixture to the leading singlet order parameter is also found. These findings  expand the physical picture of spin-split superconductivity beyond the mean-field description of the ferromagnet exchange field.
\end{abstract}

  \maketitle
 
\section{Introduction}

Long ago it was demonstrated that the exchange
field of ferromagnetic insulators (FIs), such as EuS and EuO, can spin-split the excitation spectrum of an adjacent thin-film superconductor \cite{Hao1990,Moodera1988,Tedrow1986,Meservey1994}. The spin splitting in the density of states (DOS) observed in those experiments
resembles the spin splitting created by a strong in-plane field applied to a thin
superconducting film. This discovery opened up the way for performing spin-polarized tunneling measurements without the need for applying large magnetic fields. A renewed interest in studying ferromagnetic--superconductor (F-S) structures came with active development of superconducting spintronics \cite{Linder2015,Eschrig2015}, caloritronics, and spin caloritronics \cite{Bergeret2018review,Heikkila2019review}.  
In particular, in F-S
structures with spin-split DOS, a series of promising phenomena have been studied. Among them are giant thermoelectric \cite{Machon2013,Ozaeta2014,Kolenda2016,Kolenda2016_2,Giazotto2014,Giazotto2015,Machon2014,Kolenda2017,Rezaei2018}, thermospin effects \cite{Ozaeta2014,Linder2016,Bobkova2017}, highly efficient thermally induced domain-wall motion \cite{Bobkova2021}, spin and heat valves \cite{Huertas-Hernando2002,Giazotto2006,Giazotto2008,Giazotto2013,Strambini2017}, cooling at the
nanoscale \cite{Giazotto2006review,Kawabata2013}, low-temperature thermometry and development of sensitive electron thermometers \cite{Giazotto2015_2}, detectors of electromagnetic radiation \cite{Heikkila2018,Geng2023}. 

The spin-split DOS in F-S structures has also been explored in the presence of magnetic inhomogeneities, such as textured ferromagnets and domain walls \cite{Volkov2005,Golubov2005,Ivanov2007,Strambini2017,DiBernardo2015,Diesch2018,Bobkova2019,Bobkova2019_2}.    Characteristic signatures of equal-spin-triplet pairing were reported \cite{Diesch2018}. It was shown that the characteristic spatial and energy dependence
of the spin-dependent DOS allows to tomographically extract the structure of the spin-triplet Cooper pairs \cite{Bobkova2019_2}.  Furthermore,
the influence of the domain structure on the position-averaged superconducting DOS in FI-S bilayers was studied \cite{Strambini2017}. 

Another important direction in the field of F-S hybrid structures is the investigation of interplay between the superconducting state and ferromagnetic excitations -- magnons. A series of interesting results, presumably related to the influence of the superconductor on the magnon spectrum, have been reported. In particular, it was found that the adjacent superconductor works as a spin sink strongly influencing Gilbert damping of the
magnon modes \cite{Tserkovnyak2005,Ohnuma2014,Bell2008,Jeon2019,Jeon2019_2,Jeon2019_3,Jeon2018,Yao2018,Li2018,Golovchanskiy2020,Silaev2020} and can result in shifting of $k = 0$ magnon frequencies (Kittel mode) \cite{Jeon2019_3,Li2018,Golovchanskiy2020,Silaev2020}. The electromagnetic interaction between magnons in ferromagnets and superconductors also results in the appearance of magnon-fluxon excitations \cite{Dobrovolskiy2019} and efficient gating of magnons \cite{Yu2022}. Furthermore it was reported that the magnetic proximity effect in thin-film F-S 
hybrids results in the appearance of
composite spin quasiparticles, which are 
composed of a magnon in F and an accompanying cloud
of spinful triplet pairs in S \cite{Bobkova2022}. 

Some aspects of back influence of magnons on the superconducting state have already been investigated. For example, a possible realization of the magnon-mediated superconductivity in F-S hybrids has been proposed \cite{Rohling2018,Fjaerbu2019,Erlandsen2019,Thingstad2021,Maeland2023,Gong2023}. At the same time, the influence of magnons via the magnetic proximity effect on the superconducting DOS practically has not yet been studied, although the electron-magnon interaction and influence of this interaction on the DOS in ferromagnetic metals were investigated long ago \cite{Auslender1985,Appelbaum1969}. Here we consider how the effects of electron-magnon interactions in FI-S thin-film hybrids manifest themselves in the superconducting DOS and quasiparticle spectra of the superconductor. It is found that the magnon-mediated electron spin-flip processes cause the interaction and mixing of the spin-split bands resulting in their reconstruction, which is especially important near the edge of the superconducting gap. We demonstrate  that the classical BCS-like Zeeman-split shape of the superconducting DOS can be strongly modified due to the electron-magnon interaction and this modification is temperature dependent. The influence of magnons on the temperature dependence of the Zeeman splitting of the DOS and the relevance of our findings to existing and future experiments are also discussed.

The paper is organized as follows. In Sec.~\ref{system} we describe the system under consideration and the Green's functions formalism taking into account magnon self-energies. In Sec.~\ref{spectra} the modifications of  the quasiparticle spectra in the superconductor due to the electron-magnon coupling are discussed. In Sec.~\ref{DOS} we study signatures of the electron-magnon interaction in the Zeeman-split superconducting DOS and their temperature dependence. Our conclusions are summarized in Sec.~\ref{conclusions}. Technical details of the derivation of the Green's functions are provided in Appendix A, and Appendix B is devoted to the discussion of the influence of thermal magnons on the Zeeman splitting of the DOS.   

\section{System and formalism}
\label{system}
\begin{figure}[h]
\centering
\includegraphics[width=0.4\textwidth]{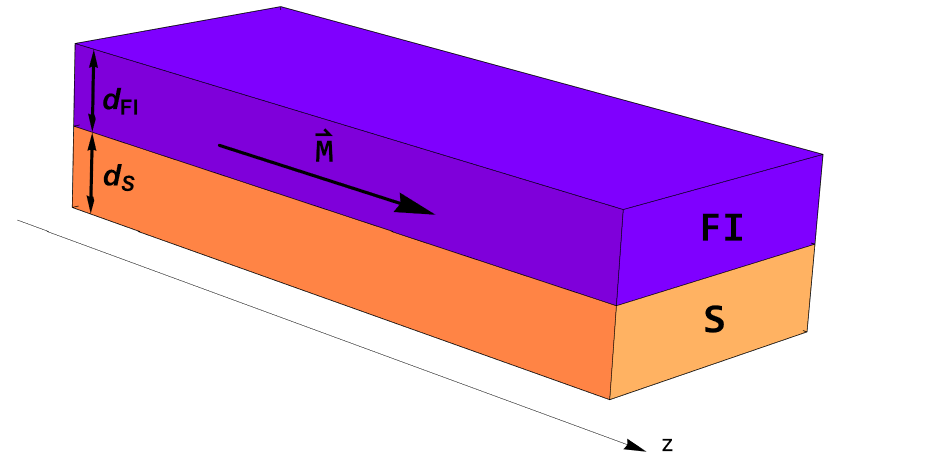}
\caption{Sketch of the FI-S thin-film bilayer.}
\label{sketch}
\end{figure}

We consider a thin-film bilayer as depicted in Fig. \ref{sketch}, in which a ferromagnetic insulator is interfaced with a conventional spin-singlet s-wave superconductor (S). The thickness of the S layer $d_S$ is assumed to be small compared to the superconducting coherence length $\xi_S$. In this case the S layer can be considered homogeneous along the normal to the interface plane. The FI layer in its ground state is magnetized in-plane, along the $z$ direction.

The Hamiltonian of the system takes the form
\begin{equation}
 \hat{H}=\hat{H}_S+\hat{H}_{FI}+\hat{H}_{ex}, 
\label{eq:ham_general}
\end{equation}
where $\hat H_S$ is the standard mean-field BCS Hamiltonian describing electrons in the superconducting film: 
\begin{equation}
    \begin{gathered}
\hat{H}_S = \sum\limits_{\bm k \sigma} \xi_{\bm k} c_{\bm k \sigma }^{\dag} c_{\bm k \sigma } -  \sum\limits_{\bm k} \Delta c_{\bm k\uparrow}^{\dag} c_{- \bm k\downarrow}^{\dag} - \sum\limits_{\bm k} \Delta^* c_{-\bm k\downarrow} c_{\bm k\uparrow} .
    \end{gathered}
\end{equation}
$\xi_{\bm k} = k^2/2m - \mu$ is the normal-state kinetic energy of the electrons in the S layer, counted from the chemical potential of the superconductor, $\mu$. $\Delta$ is the superconducting order parameter in S, which is assumed to be of conventional isotropic $s$-wave type. $c_{\bm k \sigma}^+$ and $c_{\bm k \sigma}$ are creation and annihilation operators of electrons with the wave vector  $\bm k$ and spin $\sigma$. 

$\hat H_{FI}$ describes magnons in the FI. Assuming easy-axis magnetic anisotropy in the FI it can be written as   
\begin{equation}
    \begin{gathered}
\hat{H}_{FI} = \sum\limits_{\bm q} (\omega_0 + D\bm q^2) b_{\bm q}^{\dag} b_{\bm q},  
    \end{gathered}
    \label{eq:ham_magnons}
\end{equation} 
where $b_{\bm q}^+$ and $b_{\bm q}$ are  creation and annihilation operators of magnons in the FI with wave vector $\bm q$, $\omega_0 = |\gamma| (\mu_0 H_0 + 2 K_a/M_s)$ is the magnonic frequency at $q=0$, $D$ is the  magnon stiffness constant, $\gamma$ is the typically negative gyromagnetic ratio, $M_s$ is the saturation magnetization, $\mu_0$ is the permeability of free space, $K_a$ is the easy-axis anisotropy constant and $H_0$ is the external field (can be equal to zero in our consideration).

Electronic and magnonic wave vectors $\bm k$ and $\bm q$ are assumed to be two dimensional (2D); that is, the electrons and magnons can only propagate in plane of the FI-S interface. The wave functions along the $y$ direction, perpendicular to the interface, are assumed to be quantized. For simplicity, in the formulas we leave only one transverse magnon mode. In fact, we have checked that different modes give quantitatively different, but qualitatively the same, contributions to considered self-energies. Their effect can be accounted for by multiplying our results for the self-energy corrections by an effective number of working transverse modes (see below). 

$\hat H_{ex}$ accounts for the exchange interaction between S and FI:
\begin{equation}
    \begin{gathered}
\hat{H}_{ex} = -J\int d^2 \bm \rho \bm S_{FI}(\bm \rho) \bm s_e(\bm \rho) ,
    \end{gathered}
    \label{eq:ham_coupling}
\end{equation} 
where $\bm \rho$  is a two-dimensional radius vector at  the interface plane, and $\bm S_{FI}$ and $\bm s_e$  are the spin density operators in the FI and S, respectively. $J$ is the interface exchange constant. By performing the Holstein-Primakoff transformation \cite{Kittel,HP1940,Kamra2016} to the second order in the magnonic operators in Eq.~(\ref{eq:ham_coupling}), one obtains 
\begin{equation}
    \begin{gathered}
    \label{eq:ham_ex_sum}
\hat{H}_{ex} = \hat{H}_{1} + \hat{H}_{2} + \hat{H}_{3},
    \end{gathered}
\end{equation} 
with 
\begin{equation}
    \begin{gathered}
\hat{H}_{1} = \sum\limits_{\bm k, \bm k'} U_{\bm k, \bm k'}(c_{\bm k, \uparrow}^{\dag} c_{\bm k', \uparrow}-c_{\bm k,\downarrow}^{\dag} c_{\bm k',\downarrow}) ,\\
U_{\bm k, \bm k'} = \frac{JM_s}{2|\gamma|} \int d^2 \rho \Psi_{\bm k}^*(\bm \rho) \Psi_{\bm k'}(\bm \rho),
    \end{gathered}
    \label{eq:ham_ex_1}
\end{equation} 
\begin{eqnarray}
   \hat{H}_{2} =  \sum\limits_{\bm k, \bm k',\bm q,\bm q'} T_{\bm k, \bm k',\bm q,\bm q'} b_{\bm q}^{\dag} b_{\bm q'} (c_{\bm k, \uparrow}^{\dag} c_{\bm k', \uparrow}-c_{\bm k,\downarrow}^{\dag} c_{\bm k',\downarrow}), \nonumber \\
T_{\bm k, \bm k',\bm q,\bm q'} = - \frac{J}{2} \int d^2 \rho \Psi_{\bm k}^*(\bm \rho) \Psi_{\bm k'}(\bm \rho) \phi_{\bm q}^*(\bm \rho) \phi_{\bm q'}(\bm \rho), 
    \label{eq:ham_ex_2}
\end{eqnarray} 
\begin{eqnarray}
   \hat{H}_{3} =  \sum\limits_{\bm k, \bm k', \bm q} V_{\bm k, \bm k', \bm q} (b_{\bm q} c_{\bm k, \uparrow}^{\dag} c_{\bm k', \downarrow} + b_{\bm q}^{\dag} c_{\bm k', \downarrow}^{\dag} c_{\bm k, \uparrow}), \nonumber \\
V_{\bm k, \bm k', \bm q} = J \sqrt{\frac{M_s}{2|\gamma|}} \int d^2 \rho \Psi_{\bm k}^*(\bm \rho) \Psi_{\bm k'}(\bm \rho) \phi_{\bm q}(\bm \rho) ,
        \label{eq:ham_ex_3} 
\end{eqnarray} 
where $\hat H_1$ describes a spin splitting of
the electronic energy spectrum in S in the mean-field approximation. The second term $\hat H_2$ represents the Ising term, which physically accounts for the renormalization of the spin splitting by magnonic contribution. Since the processes of the spin transfer between electrons and magnons are of primary importance for our consideration, when calculating the electronic Green's function we simplify this term by substituting the magnon operator $b_{\bm q}^\dagger b_{\bm q}$ by its averaged value $\langle b_{\bm q}^\dagger b_{\bm q}\rangle = n_{\bm q} \delta_{\bm q \bm q'}$, where $n_{\bm q}$ is the density of magnons with wave vector $\bm q$. The third term $\hat H_3$ transfers spin between electron and magnon operators and will turn out to be the most significant for the effects under consideration. 

If we choose the wave functions of electrons $\Psi_{\bm k}(\bm \rho)$ and magnons $\phi_{\bm q}(\bm \rho)$ at the interface in the form of plane waves propagating along the interface, that is, $\Psi_{\bm k}(\bm \rho)=(1/\sqrt{d_S})e^{i \bm k \bm \rho}$ and $\phi_{\bm q}(\bm \rho)=(1/\sqrt{d_{FI}})e^{i \bm q \bm \rho}$, then $\hat{H}_{ex}$ can be simplified: 
\begin{equation}
    \begin{gathered}
\hat{H}_{ex} = \tilde U \sum\limits_{\bm k} (c_{\bm k, \uparrow}^{\dag} c_{\bm k, \uparrow}-c_{\bm k,\downarrow}^{\dag} c_{\bm k,\downarrow}) + \\  V \sum\limits_{\bm k, \bm q} (b_{\bm q} c_{\bm k, \uparrow}^{\dag} c_{\bm k-\bm q, \downarrow} + b_{\bm q}^{\dag} c_{\bm k-\bm q, \downarrow}^{\dag} c_{\bm k, \uparrow}) ,
    \end{gathered}
\end{equation} 
where $\tilde U = -J (M_s-N_m |\gamma|)/(2|\gamma|d_S )$ is the averaged spin-splitting field in the superconductor renormalized by the magnon density $N_m$, and $V = J\sqrt{M_s/2|\gamma|d_{FI} A}(1/d_S)$ is the electron-magnon coupling constant, where $A$ is the area of the FI-S interface.

Introducing the following Nambu spinor $\check \Psi_{\bm k} = (c_{\bm k \uparrow}, c_{\bm k \downarrow}, -c_{-\bm k \downarrow}^\dagger, c_{-\bm k \uparrow}^\dagger)^T$, we define the Gor'kov Green's function in the Matsubara representation, $\check G_{\bm k}(\tau) = -\langle T_\tau \check \Psi_{\bm k} \check \Psi_{\bm k}^\dagger \rangle$, where $\langle T_\tau ... \rangle$ means imaginary time-ordered thermal averaging. Turning to the Matsubara frequency representation, the Green's function obeys the following equation:
\begin{eqnarray}
(i\omega  - \xi_k \tau_z - \tilde U \sigma_z - \Delta \tau_x - \check{\Sigma}_m )\check{{G}}_{\bm k } (\omega) = 1, 
\label{eq:Gorkov} 
\end{eqnarray}
where $\omega$ is the fermionic Matsubara frequency, and $\sigma_i$ and $\tau_i$ ($i=x,y,z$) are Pauli matrices in spin and particle-hole spaces, respectively. $\check \Sigma_m$ is the magnonic self-energy, which describes corrections to the electronic Green's function due to the electron-magnon interaction and in the framework of the self-consistent Born approximation takes the form:
\begin{eqnarray}
 \check{\Sigma}_m = - V^2 T \sum \limits_{\bm q,\Omega} B_{\bm q}(\Omega) \left\{\sigma_+ \check{G}_{\bm k-\bm q} (\omega - \Omega)\sigma_- + \right. \nonumber \\
 \left. \sigma_- \check{G}_{\bm k+\bm q} (\omega +  \Omega)\sigma_+\right\} , 
    \label{eq:magnon_se}
\end{eqnarray} 
where $\sigma_{\pm} = (\sigma_x \pm i \sigma_y)$, $\Omega$ is the bosonic Matsubara frequency, and  $B_{\bm q}(\Omega) = [i\Omega - (\omega_0+Dq^2)]^{-1}$ is the magnonic Green's function. From the spin structure of Eq.~(\ref{eq:magnon_se}) it is seen that $\check \Sigma_m$ is diagonal in spin space. For this reason the electronic Green's function, which is given by the solution of Eq.~(\ref{eq:Gorkov}) is also a diagonal matrix in spin space and Eq.~(\ref{eq:Gorkov}) can be written for the both spin subbands separately:
\begin{eqnarray}
(i\omega  - \xi_k \tau_z - \sigma \tilde U  - \Delta \tau_x - \hat{\Sigma}_{m, \sigma
} )\hat{{G}}_{\bm k, \sigma } (\omega) = 1, 
\label{eq:gorkov_spin}
\end{eqnarray}
where $\hat G_{\bm k, \sigma}$ is $2 \times 2$ matrix in the particle-hole space corresponding to the electron spin $\sigma = \uparrow, \downarrow$. $\hat \Sigma_{m,\sigma}$ is also a $2 \times 2$ matrix in the particle-hole space representing the magnonic self-energy for the given spin subband $\sigma$:
\begin{eqnarray}
 \hat{\Sigma}_{m,\sigma} = - V^2 T \sum \limits_{\bm q,\Omega} B_{\bm q}(\Omega) \hat {G}_{\bm k-\sigma\bm q, \bar \sigma} (\omega - \sigma \Omega), 
    \label{eq:magnon_se_spin}
\end{eqnarray} 
where a factor $\sigma$ means $\pm 1$ for the spin-up (spin-down) subbands, and $\bar \sigma$ means the opposite spin subband. The dimensionless coupling constant quantifying the strength of the electron-magnon coupling is $K=V^2 A / 4 \pi \hbar v_F \sqrt{D \Delta}$. Our numerical estimates made for the parameters corresponding to EuS-Al or YIG-Nb  structures suggest that K should be rather small, $K \ll 1$ (for a detailed discussion of the numerical
estimates, see Sec.~\ref{DOS}). The smallness of the electron-magnon coupling constant allows us to use a non-self-consistent Born approximation when calculating magnon self-energy. That is, we substitute $\hat G_{\bm k - \sigma \bm q, \bar \sigma}$ by the bare superconducting Green's function obtained without taking into account the magnon self-energy $\hat G_{\bm k - \sigma \bm q, \bar \sigma}^{(0)}$ in Eq.~(\ref{eq:magnon_se_spin}). Then the explicit solution of Eq.~(\ref{eq:gorkov_spin}) takes the form
\begin{equation}
    \begin{gathered}
\hat{{G}}_{\bm k,\sigma} (\omega) = \frac{i \widetilde{\omega}_{\bm k, \sigma} +\widetilde{\xi}_{\bm k, \sigma} \tau_z + \widetilde{\Delta}_{\bm k, \sigma} \tau_x}{(i \widetilde{\omega}_{\bm k, \sigma})^2 - (\widetilde{\xi}_{\bm k, \sigma})^2 - (\widetilde{\Delta}_{\bm k, \sigma})^2} ,
    \end{gathered}
    \label{eq:gorkov_sol}
\end{equation} 
where all the quantities marked by $\tilde ~$ are renormalized by the electron-magnon interaction as follows (see Appendix A for a more detailed derivation of the magnonic corrections):
\begin{equation}
    \begin{gathered}
\widetilde{\Delta}_{\bm k, \sigma} (\omega) = \Delta + \delta \Delta_{\bm k,\sigma}(\omega) = \Delta - \\ - V^2 T \sum\limits_{\bm q, \Omega} B_{\bm q}(\Omega) \frac{\Delta}{(i \omega - i\sigma \Omega +\widetilde{U} \sigma)^2 - \xi^2_{\bm k-\sigma \bm q} - \Delta^2} ,
    \end{gathered}
    \label{eq:delta_corr}
\end{equation} 
\begin{equation}
    \begin{gathered}
\widetilde{\xi}_{\bm k, \sigma} (\omega) = \xi_{\bm k} + \delta \xi_{\bm k,\sigma}(\omega)= \xi_{\bm k} -  \\ - V^2 T \sum\limits_{\bm q, \Omega} B_{\bm q}(\Omega) \frac{\xi_{\bm k-\sigma \bm q}}{(i \omega - i\sigma \Omega +\widetilde{U} \sigma)^2 - \xi^2_{\bm k-\sigma \bm q} - \Delta^2} ,
    \end{gathered}
    \label{eq:xi_corr}
\end{equation}
\begin{equation}
    \begin{gathered}
i\widetilde {\omega}_{\bm k, \sigma} (\omega) = i \omega - \widetilde{U} \sigma + i\delta \omega_{\bm k,\sigma}(\omega)= i \omega - \widetilde{U} \sigma + \\ +  V^2 T \sum\limits_{\bm q, \Omega} B_{\bm q}(\Omega) \frac{i \omega - i\sigma \Omega +\widetilde{U} \sigma}{(i \omega - i\sigma \Omega +\widetilde{U} \sigma)^2 - \xi^2_{\bm k-\sigma \bm q} - \Delta^2} .
    \end{gathered}
    \label{eq:energy_corr}
\end{equation}

\begin{figure}[tb]
\begin{center}
\includegraphics[width=80mm]{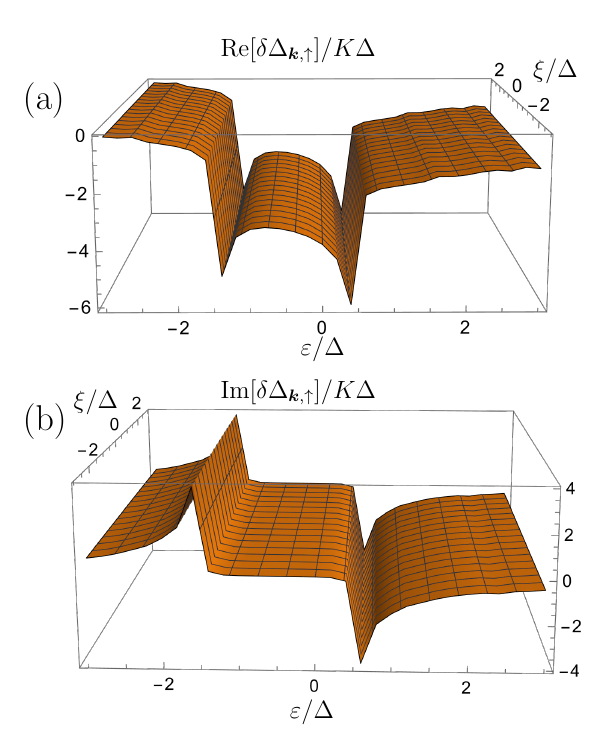}
\caption{(a) ${\rm Re}[\delta \Delta_{\bm k,\uparrow}]/K\Delta$ and (b) ${\rm Im}[\delta \Delta_{\bm k,\uparrow}]/K\Delta$ as functions of $\xi$ and quasiparticle energy $\varepsilon$. $K$ is the dimensionless quantity characterizing the electron-magnon coupling strength (see text for its exact definition). Here and below throughout the calculations we take $\omega_0=0.03\Delta$,
$\delta=0.01\Delta$, and
$D_m/\xi_S^2=10^{-4}\Delta$. 
}
\label{fig:delta_corr}
\end{center}
\end{figure}

\begin{figure}[tb]
\begin{center}
\includegraphics[width=80mm]{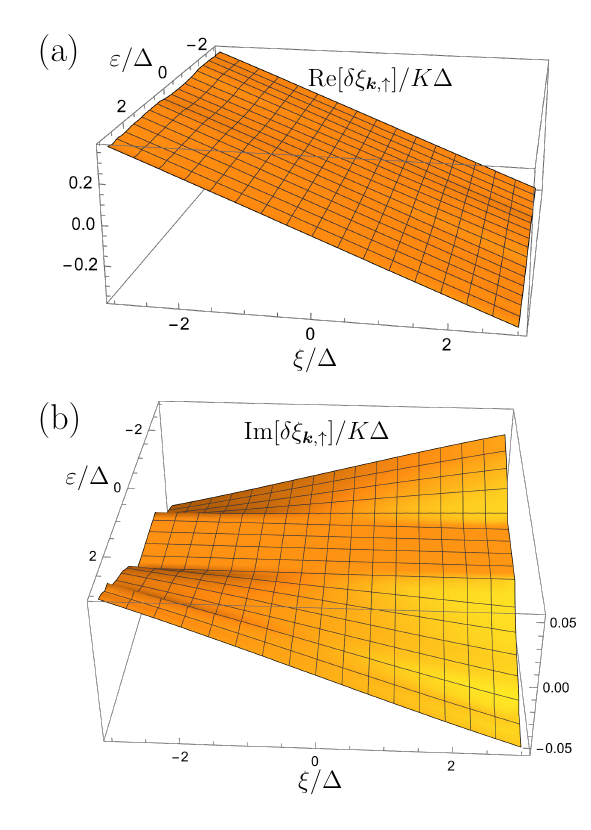}
\caption{(a) ${\rm Re}[\delta \xi_{\bm k,\uparrow}]/K\Delta$ and (b) ${\rm Im}[\delta \xi_{\bm k,\uparrow}]/K\Delta$ as functions of $\xi$ and  $\varepsilon$.}
\label{fig:xi_corr}
\end{center}
\end{figure}

\begin{figure}[tb]
\begin{center}
\includegraphics[width=80mm]{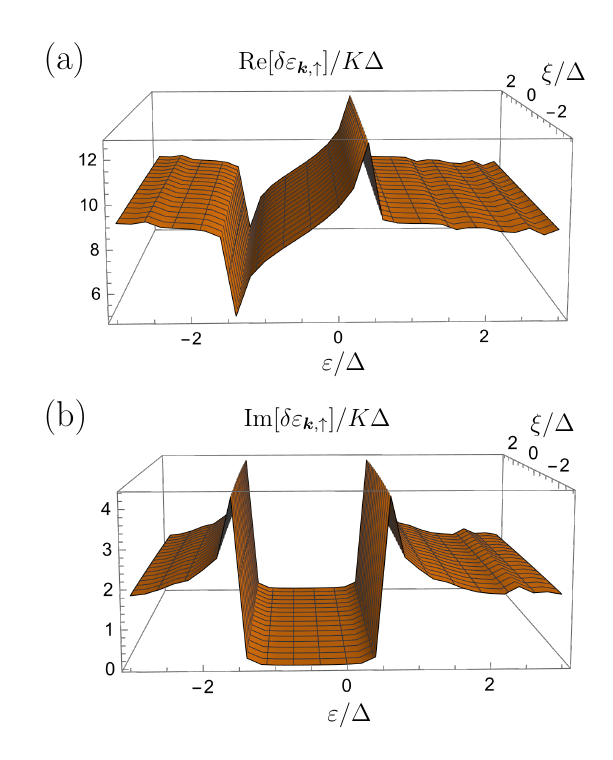}
\caption{(a) ${\rm Re}[\delta \varepsilon_{\bm k,\uparrow}]/K\Delta$ and (b) ${\rm Im}[\delta \varepsilon_{\bm k,\uparrow}]/K\Delta$ as functions of $\xi$ and  $\varepsilon$.}
\label{fig:energy_corr}
\end{center}
\end{figure}

For the problem under consideration all the in-plane directions of $\bm k$ are equivalent. For this reason the magnonic corrections only depend on the absolute value $k$ of the wave vector. Furthermore, in order to study the quasiparticle spectra and density of states we turn from Matsubara frequencies to the real energies in the Green's functions, $i \omega \to \varepsilon + i \delta$, where $\delta$ is an infinitesimal positive number.  The magnonic corrections for spin-up electrons $\delta \Delta_{\bm k, \uparrow}$, $\delta \xi_{\bm k, \uparrow}$, and $\delta \varepsilon_{\bm k, \uparrow} = i\delta \omega_{\bm k, \uparrow}(i\omega \to \varepsilon + i \delta)$ are presented in Figs.~\ref{fig:delta_corr}-\ref{fig:energy_corr}  as functions of the quasiparticle energy $\varepsilon$ and $\xi_{\bm k} \equiv \xi$, which after linearization in the vicinity of the Fermi surface takes the form $\xi_{\bm k } \approx \bm {v}_F (\bm k - \bm k_F)$.

The key features of the corrections, which can be seen in the presented plots, are as
follows:

(i) The dependence of the corrections on $\xi$ is very weak. The reason is that the most important range of magnonic wave numbers contributing to the corrections is $q \lesssim 1/\xi_S$, where $\xi_S = v_F/\Delta$ is the superconducting coherence length. Then taking parameters of the magnon spectrum corresponding to YIG  $\omega_{0,YIG} = 3 * 10^{-2}\Delta $ $D_{YIG} \approx 5*10^{-40}J*m^2$ or EuS, $\omega_{0,EuS} \sim 10^{-2}\Delta$, $D_{EuS} \approx 3*10^{-42}J*m^2 $, we obtain that $D q^2 \ll \omega_0$ to very good accuracy for all reasonable parameters. Consequently, one can disregard $D q^2$ with respect to $\omega_0$ in the magnonic Green's function $B_{\bm q}$ and after linearization of $\xi_{\bm k - \sigma \bm q} \approx \bm {v}_F (\bm k - \sigma \bm q - \bm k_F)$ in the vicinity of the Fermi surface we see that the dependence on $\bm k$ drops from Eqs.~(\ref{eq:delta_corr})-(\ref{eq:energy_corr}).

(ii)
The correction to the normal-state electron dispersion $\delta \xi$ is small with respect to all other corrections and is neglected below.

(iii) The important corrections $\delta \Delta$ and $\delta \varepsilon$ have peaks at the energies corresponding to the superconducting coherence peaks of the {\it opposite} spin subbands. While the coherence peaks for the spin-up subband are located at $\varepsilon = \pm \Delta +\tilde U$, the peaks of the corrections are at $\varepsilon = \pm \Delta -\tilde U$. It is an obvious consequence of the process of electron spin flip accompanied by emission or absorption of a magnon.

(iv) The correction $\delta \Delta$ represents an effective contribution to the superconducting order parameter induced from the pure singlet pairing $\Delta$ via the electron-magnon interaction. It depends on the Matsubara frequency and contains both singlet and triplet components. As can be seen from Eq.~(\ref{eq:delta_corr}), the correction  obeys the condition $\delta \Delta_{\uparrow}(\omega)$ = $\delta \Delta_{\downarrow}(-\omega)$. It means that the triplet component $\delta \Delta_t (\omega) = \delta \Delta_{\uparrow}(\omega) - \delta \Delta_{\downarrow}(\omega) = -\delta \Delta_t(-\omega)$ works as an effective odd-frequency {\it superconducting order parameter}. This situation is rather unusual because typically in F-S hybrid systems we encounter an odd-frequency anomalous Green's function, but at the same time the order parameter is still even frequency in the framework of the conventional BCS weak coupling theory.

\section{Quasiparticle spectra}
\label{spectra}

Now we turn to a discussion of how quasiparticle spectra in the S layer are modified by the electron-magnon interaction.
\begin{figure}[tb]
\begin{center}
\includegraphics[width=90mm]{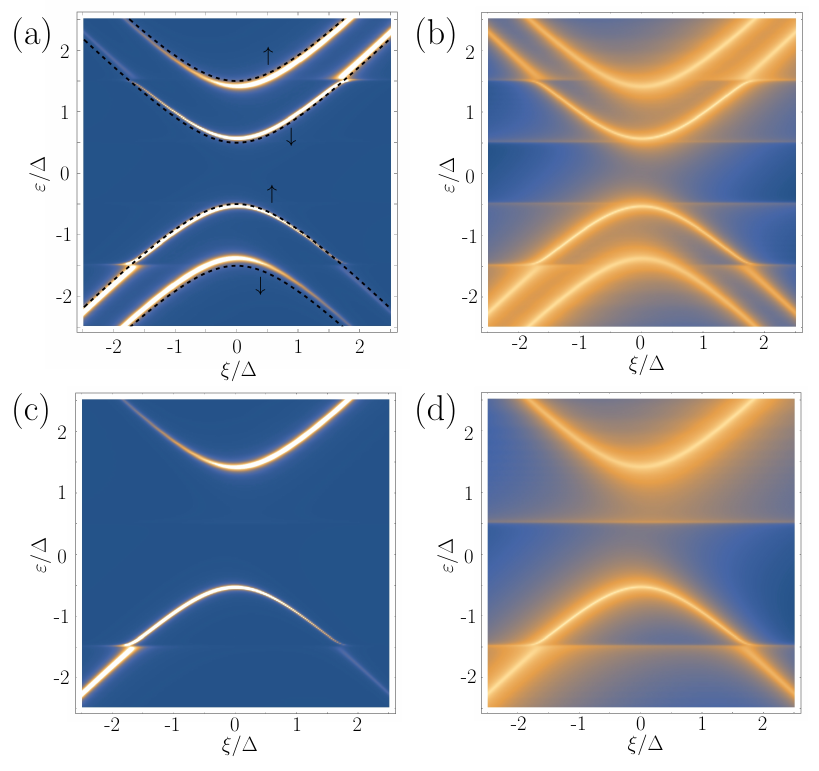}
\caption{(a) Electronic spectral function in the S layer for both spins  $A_{\uparrow,\downarrow}(\varepsilon, \xi)$ at $K=0.01$. Dashed lines represent the spin-up and spin-down  branches of the spectrum without taking into account the electron-magnon interaction, that is, at $K=0$. $T = 0.1 \Delta$, $\tilde U=0.5 \Delta$. (b) The same on a logarithmic scale. (c) Only spin-up electronic spectral function in the S layer $A_{\uparrow}(\varepsilon, \xi)$ at $K=0.01$. The spin-down spectral function can be found as $A_{\downarrow}(\varepsilon, \xi) = A_{\uparrow}(-\varepsilon, -\xi)$. (d) The same as in (c) on a logarithmic scale.}
\label{fig:spectra}
\end{center}
\end{figure}
In Fig.~\ref{fig:spectra}(a) we present the spectral functions for both spins in the S layer calculated from the Green's function (\ref{eq:gorkov_sol}) according to the relation
\begin{eqnarray}
A_\sigma(\varepsilon, \bm k) = -\frac{1}{\pi}{\rm Tr}\left\{\frac{1+\tau_z}{2}{\rm Im}[\hat G_{\bm k,\sigma}^R(\varepsilon)]\right\}.
\label{eq:spectral_function}
\end{eqnarray}

\begin{figure}[tb]
\begin{center}
\includegraphics[width=90mm]{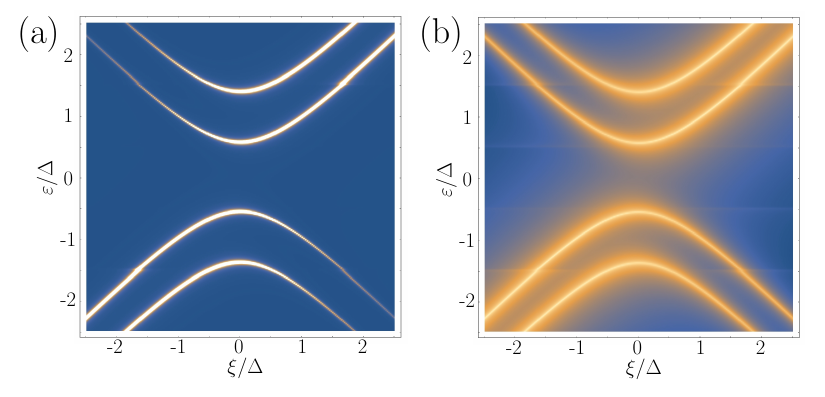}
\caption{The same as in Figs.~\ref{fig:spectra}(a)-(b), but for lower temperature $T = 0.01 \Delta$. The other parameters are the same. The reconstruction and broadening of the spectra are much less pronounced.}
\label{fig:spectraT001}
\end{center}
\end{figure}

The spectral function is isotropic in momentum space and for this reason we plot it as a function of $\xi_{\bm k} \equiv \xi$. The electron like and hole like quasiparticle branches are clearly seen at positive and negative energies, respectively. Black dashed lines represent the quasiparticle spectra in the absence of the electron-magnon interaction. The electron-magnon interaction leads to the following main modifications of the quasiparticle spectra:

(i) The Zeeman splitting of spin-up and spin-down quasiparticle branches is reduced due to the magnon-mediated interaction between quasiparticles with opposite spins.

(ii) For positive energy branches, corresponding to electron-like quasiparticles, the lifetime of spin-up  quasiparticles and quasiparticles at the upper part of the spin-down branch is considerably suppressed, which is seen as a broadening of the corresponding branches. For negative energies, corresponding to hole like quasiparticles, the situation is symmetric if we interchange spins. The broadening of the spin-down branch only occurs in the energy region, where the spin-up branch also exists. The physical reason is that the spin-flip processes providing the broadening are nearly horizontal due to the fact that $\omega_0 + Dq^2 \ll \Delta$; that is, the magnon energies are small as compared to $\Delta$ in the whole range of $\xi$, considered in Fig.~(\ref{fig:spectra}). The lower (upper) part of the spin-down (up) positive (negative) energy branch is not broadened because there are no available states for the opposite-spin quasiparticles at the appropriate energies and, consequently, the spin-flip processes are not allowed.

(iii) In Fig.~\ref{fig:spectra}(a) we also see a reconstruction of the spin-down spectral branch in the energy range of the bottom of the spin-up branch. In order to investigate this effect in more detail we plot the same figure on a logarithmic scale in Fig.~\ref{fig:spectra}(b), which allows to clearly see weak spectral features. Figures ~\ref{fig:spectra}(c) and ~\ref{fig:spectra}(d) represent the spectral functions for the spin-up band on the normal and on the logarithmic scale, respectively. From Figs.~\ref{fig:spectra}(b) and ~\ref{fig:spectra}(d) it is seen that due to the electron-magnon interaction in the energy region of the extremum of the spin-up (down) branch, a nonzero density of states appears for the opposite-spin branch. It looks like a horizontal line starting from the bottom of the corresponding branch. This line is horizontal due to the independence of the electron-magnon self-energy corrections (\ref{eq:delta_corr}) and (\ref{eq:energy_corr}) on $\xi$. This mixing of the spin-up and spin-down bands resulting from the magnon-mediated spin-flip processes is natural and exists at all energies, but the spectral weight of the opposite-spin branch is too small except for the regions of the extrema of the bands corresponding to the coherence peaks of the superconducting DOS. Intersection of the additional lines with the original spin-down band results in its reconstruction, which looks like an avoided crossing point.

The results for the spectral function presented and discussed above correspond to $T=0.1\Delta$. This temperature is higher than the gap in the magnonic spectrum, $\omega_0=0.03\Delta$, which we take in our calculations. Therefore, a large number of thermal magnons are excited at this temperature. In Fig.~\ref{fig:spectraT001} the spectral function is demonstrated for lower temperature, $T=0.01\Delta<\omega_0$. It is seen that the characteristic signatures of the magnon-mediated spin-flip processes,  that is, the mixing, reconstruction, and broadening of the branches, are much less pronounced due to the suppression of the thermally excited magnons at such low temperatures.

\section{DOS in the presence of magnons}
\label{DOS}

\begin{figure}[tb]
\begin{center}
\includegraphics[width=85mm]{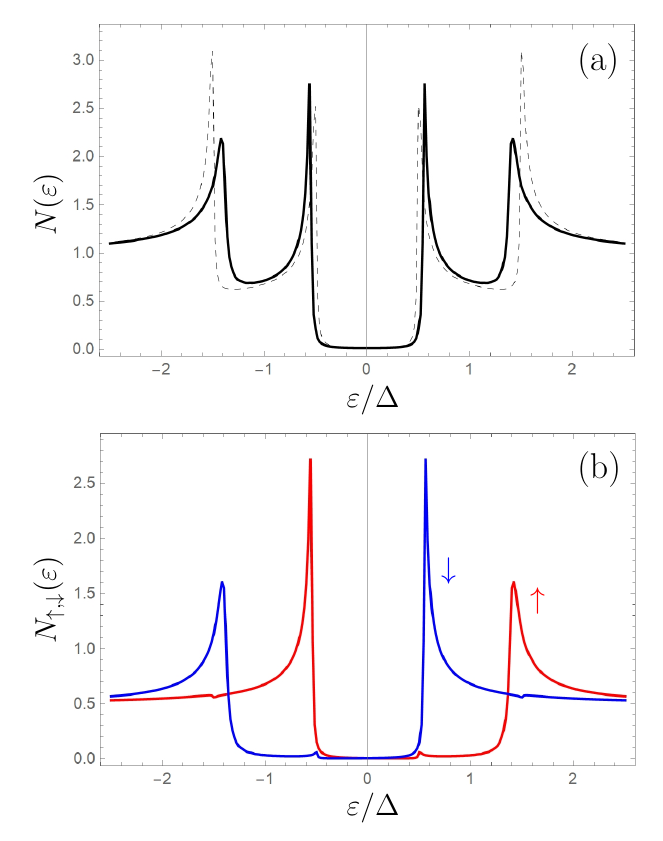}
\caption{(a) LDOS in the S layer with (solid line, $K=0.01$) and without (dashed, $K=0$) taking into account electron-magnon interaction. (b) Spin-resolved LDOS $N_{\uparrow}$ (red) and $N_\downarrow$ (blue) for $K=0.01$. The solid line in panel (a) is obtained by summing red and blue curves from panel (b). $T=0.1\Delta$, $\tilde U = 0.5 \Delta$.}
\label{fig:LDOS}
\end{center}
\end{figure}

Now we turn to a discussion of the local density of states (LDOS) in the S layer, which is calculated as the momentum-integrated spectral function:
\begin{eqnarray}
N(\varepsilon) = \int \frac{d^2k}{(2\pi)^2} A(\varepsilon,\bm k).
\label{eq:ldos}
\end{eqnarray}
Figure~\ref{fig:LDOS}(a) demonstrates the LDOS in the presence of electron-magnon interaction (solid line) as compared to the LDOS calculated at $V=0$ (dashed line). The LDOS at $V=0$, which is calculated assuming mean-field approximation for the exchange field, takes the conventional BCS-like shape. It manifests Zeeman-split coherence peaks, and the outer peak is always higher than the inner one. This BCS-like shape is typical for thin-film FI-S bilayers with $d_S \ll \xi_S$ and thin superconducting films in the applied parallel magnetic field \cite{Hao1990,Moodera1988,Tedrow1986,Kolenda2016,Strambini2017}. For thicker superconducting films with $d_S \gtrsim \xi_S$ the shape of the DOS is modified and becomes  dependent on the coordinate along the normal to the FI-S plane \cite{Hijano2021}. Although the Zeeman splitting is still present in the vicinity of the FI-S interface, the coherence peaks can be smeared and, therefore, the manifestations of the electron-magnon interaction in the DOS in this case is a separate task, which is beyond the scope of the present work. The electron-magnon interaction can invert the relative ratio of the peak heights and broadens the outer peaks, while the width of the inner peaks remains unchanged. The reason is the same as for the broadening of the spectra in Fig.~\ref{fig:spectra}: electron spin-flip processes accompanied by a magnon emission or absorption. The outer coherence peaks in Fig.~\ref{fig:LDOS}(a) correspond to the energy regions of the bottom (top) of the positive(negative)-energy spin-up(down) bands. This type of broadening, which only affects outer peaks, differs from the other physical mechanisms resulting in the broadening of the coherence peaks, such as the orbital effect of the magnetic field, inelastic scattering or magnetic impurities, which affect all the peaks \cite{Meservey1994} and can be roughly described by the Dynes parameter. 

Figure~\ref{fig:LDOS}(b) represents the spin-resolved LDOS $N_\uparrow$ (red) and $N_\downarrow$
(blue). The solid line in Fig.~\ref{fig:LDOS}(a) is obtained
by summing red and blue curves from Fig.~\ref{fig:LDOS}(b). Figure~\ref{fig:LDOS}(b) additionally illustrates the unique magnon-induced mechanism of broadening of the outer peaks of the LDOS. As discussed in Sec.~\ref{spectra}, the
broadening of the spin-down branch only occurs in the
energy region where the spin-up branch also exists. The same is valid for the spin-up branch. Physically it is a signature of nearly horizontal spin-flip processes accompanied by a magnon emission or absorption. In order to have a possibility to flip the spin, one needs an available space in the opposite-spin subband at the corresponding energy.  This broadening of the quasiparticle branches manifests itself as an asymmetric broadening of coherence peaks in Fig.~\ref{fig:LDOS}(b). Only one of the coherence peaks on the blue curve, which corresponds to the energy, where we have a nonzero DOS in the red subband, is broadened. The same is valid for the red subband. 

\begin{figure}[tb]
\begin{center}
\includegraphics[width=85mm]{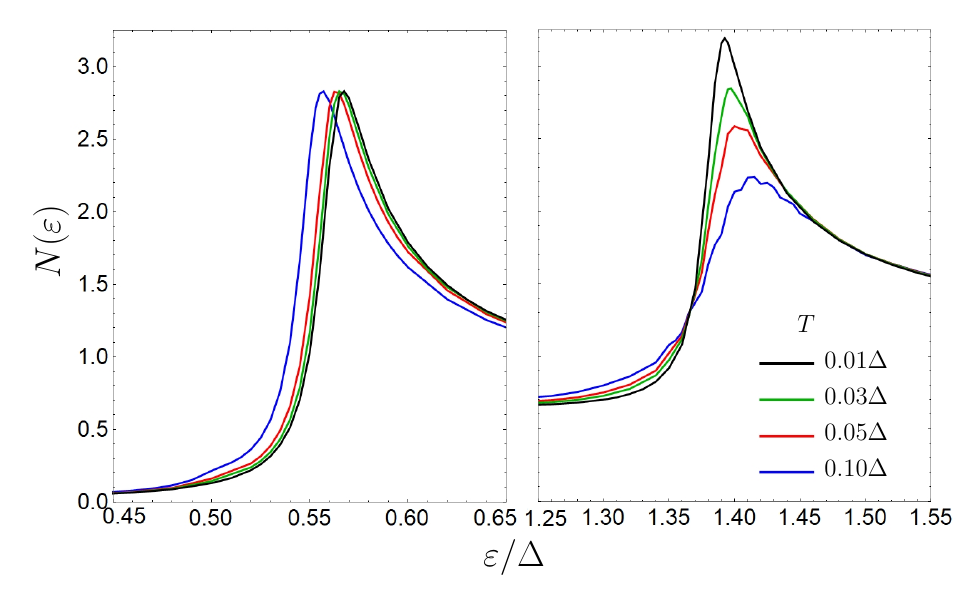}
\caption{Inner LDOS peak (left) and outer LDOS peak (right) as functions of energy. Different curves correspond to different temperatures. $K=0.01$, $\tilde U = 0.5 \Delta$.}
\label{fig:LDOS_temp}
\end{center}
\end{figure}

The other important manifestation of the electron-magnon interaction is that the shape of the LDOS strongly depends on temperature even at very low temperatures $\sim \omega_0 \ll \Delta$, in agreement with the above-discussed behavior of the spectral function. The temperature evolution of the LDOS is presented in Fig.~\ref{fig:LDOS_temp}. It is seen that the broadening of the outer peak develops with increasing temperature in the temperature range $\sim \omega_0$. It is clear if we remember that the broadening is caused by the spin-flip processes, which are mediated by the thermally excited magnons. We do not consider larger temperatures $T \gg \omega_0 $ comparable to the critical temperature of the superconducting film because in this temperature range the temperature dependence of the superconducting gap comes into play and correct consideration of the problem requires solution of a self-consistency equation for the order parameter.

Now let us discuss numerical estimates of the dimensionless constant $K=V^2 A / 4 \pi \hbar v_F \sqrt{D \Delta}$, which controls the strength of the electron-magnon coupling. Substituting $V = J\sqrt{M_s/2|\gamma|d_{FI} A}(1/d_S)$ and expressing the interface exchange coupling constant via the experimentally accessible quantity $\tilde U$ as $|J| = 2 |\gamma| \tilde U d_S/M_s$ (where to the leading approximation we neglect magnonic contribution to the magnetization), we obtain $K = \tilde U^2 (2|\gamma|/M_s) 1/(4 \pi \sqrt{D\Delta} \hbar v_F d_{FI})$ for one transverse magnon mode. The effective number of working transverse modes $N_\perp \sim d_{FI}/a$, where $a$ is the interatomic distance in the ferromagnet. According to our estimates for $d_{FI} \approx 10 nm$ $N_\perp \sim 2 \div 5$. One can take the following parameters for YIG-Nb heterostructures: $\tilde U/\Delta = 0.5$ ($\tilde U$ was estimated in \cite{Kamra2018} as $1 \div 10$T, which corresponds to $h \sim 10^{-23} \div 10^{-22}$ J), $v_F = 10^6m/s$, $\Delta_{Nb} = 2.7*10^{-22}J$, $a=1.2nm$, $2|\gamma|/M_s = 3.3*10^{-27}m^3$, $D = D_{bare,YIG}-\delta D_{YIG}$, where $D_{bare,YIG} = 5*10^{-40}J*m^2$\cite{Xiao2010} is the exchange stiffness of $YIG$ and $\delta D_{YIG}$ is the renormalization of the stiffness in FI-S bilayers due to the formation of composite spin quasiparticles \cite{Bobkova2022}. As it was predicted \cite{Bobkova2022}, for the material parameters of YIG-Nb heterostructures $\delta D_{YIG}$ can be $\sim (0.5 \div 1) D_{YIG,bare}$ for $d_{FI} \sim (1 \div 0.5) d_S$. Therefore, the electron-magnon coupling constant for YIG-Nb heterostructures can vary in the wide range $K_{YIG-Nb} \gtrsim 10^{-4}$. The values considered here, $K \sim 0.01$, can be realized in the regime of strong renormalization of the exchange stiffness constant $D$. It is also worth noting that there is rather strong impurity-induced spin-orbit interaction in  Nb. It is known \cite{Heikkila2019review} that the spin-orbit scattering mixes both spin subbands thus killing Zeeman splitting of the coherence peaks at large values of the parameter $(\tau_{so} T_c)^{-1} \gg 1$, where $\tau_{so}^{-1}$ is the strength of the spin-orbit scattering and $T_c$ is the critical temperature of the superconductor.  It was reported \cite{Wakamura2014} that at low temperatures for Nb $\tau_{so} \approx 10^{-12}$s, which means $(\tau_{so} T_c)^{-1} \approx 1$. For this value of spin-orbit strength the Zeeman splitting of the superconducting DOS can be clearly resolved \cite{Heikkila2019review}.

For EuS-Al heterostructures one can take $\tilde U/\Delta = 0.25$ \cite{Strambini2017}, $v_F = 10^6m/s$, $\Delta_{Al} = 3.5*10^{-23}J$, $a=10^{-10}m$, $2|\gamma|/M_s = 3.3*10^{-28}m^3$, $D = D_{bare,EuS}$, where $D_{bare,EuS} = 3*10^{-42}J*m^2$\cite{Boni1995}. The superconducting renormalization of the stiffness due to the formation of composite spin quasiparticles is predicted to be small for the parameters corresponding to EuS-Al heterostructures at reasonable thicknesses $d_{FI}$ due to smaller values of $\Delta$ and larger $M_s$. Substituting these parameters into the expression for $K$ we come to the conclusion that for EuS-Al heterostructures $K_{EuS-Al} \sim 10^{-7} \div 10^{-6}$; that is, the electron-magnon effects are unlikely to be observed in such structures. 

In general, the electron-magnon effects in the LDOS and quasiparticle spectra should be more pronounced in ultrathin superconducting films with high critical temperatures, where large absolute values of the effective exchange field $\tilde U$ can be realized. The smaller values of the exchange stiffness of the ferromagnet will also enhance the effect. The manifestations of the electron-magnon coupling become more pronounced at $T \gtrsim \omega_0$ and grow with temperature. 

Now we discuss the influence of the electron-magnon interaction on the effective Zeeman splitting, which is defined as the distance between the split coherence peaks of the LDOS divided by 2.  Experimentally, the low-temperature reduction of the effective Zeeman splitting at $T \ll \Delta$ for EuS-Al heterostructures has been reported \cite{Strambini2017}. It was ascribed to the presence of weakly bound spins at the EuS-Al interface. The renormalization of the effective exchange field in the superconductor by the thermal magnons can also contribute to this effect. Indeed, the fit of experimentally observed temperature dependence of the distance between the Zeeman-split coherence peaks $\Delta V_{peaks}(T)$ by $2|\tilde U| = J (M_s-N_m |\gamma|)/(|\gamma|d_S )$  with the magnon density $N_m = (1/S d_{FI})\sum_{\bm q}\left\{\exp[-(\omega_0+Dq^2)/T]-1\right\}^{-1}$ and $\omega_0 \approx 0.03K$ is in reasonable agreement with the experimental data and is presented in Appendix B. 

In addition, the broadening of the outer coherence peaks, predicted in this work, leads to  enhancement of the distance between the spin-split coherence peaks. The broadening becomes stronger with increasing temperature. This effect leads to an apparent growth of the peaks splitting with temperature and, therefore, acts opposite to the renormalization of the effective Zeeman field by magnons. However, our numerical estimates suggest that the temperature growth is unlikely to be observed, at least for heterostructures, consisting of the materials discussed above, because the renormalization of the effective Zeeman field by magnons dominates.

\section{Conclusions}
\label{conclusions}

In this work the influence of the electron-magnon interaction at the superconductor--ferromagnetic-insulator interface in thin-film FI-S heterostructures on the spectrum of quasiparticles and the LDOS in the superconducting layer is studied. It is predicted that due to the magnon-mediated electron spin-flip processes the spin-split quasiparticle branches are partially mixed and reconstructed. The reconstruction is the most pronounced in the region of the bottom of the energetically unfavorable spin band because of the enhanced density of the electronic states and existence of the available states in the opposite-spin band. The BCS-like Zeeman-split shape of the superconducting DOS, which is typical for superconductors in the effective exchange field,  is strongly modified due to the electron-magnon interaction. The outer spin-split coherence peaks are broadened, and the inner peaks remain intact. This type of broadening is a clear signature of the magnon-mediated spin flips and strongly differs from other mechanisms of the coherence peak broadening, which usually influence all peaks.  The broadening grows with temperature due to the thermal excitation of magnons. The above-described features in the electronic DOS are mainly caused by magnonic contributions that are diagonal
in the particle-hole space to the electron self-energy, that is, by the quasiparticle processes. Besides that we have also
found a magnonic contribution that is off-diagonal in the particle-hole space to the electronic self-energy. It mimics an odd-frequency superconducting order parameter admixture to the leading singlet order parameter. The study of its influence on the superconducting properties of the system may be an interesting direction for future research.

\section{Acknowledgments}
We acknowledge the discussions of the exchange interaction Hamiltonian with Akashdeep Kamra. The work was supported by the Russian Science Foundation via the RSF Project No.~22-42-04408. 

\section*{Appendix}

\subsection{Derivation of the expression for the Green's function and magnonic self-energy corrections}

As described in Sec.~\ref{system}, the electronic Green's function is a diagonal matrix in spin space and is to be found from Eq.~(\ref{eq:gorkov_spin}), where the magnon self-energy for a  given spin subband $\sigma$ is expressed by Eq.~(\ref{eq:magnon_se_spin}). The magnon self-energy is a $2 \times 2$ matrix in particle-hole space and in general can be expanded over Pauli matrices $\tau_i$ as follows:
\begin{eqnarray}
\hat \Sigma_{m,\sigma} = -i\delta \omega_{\bm k,\sigma} + \delta \xi_{\bm k,\sigma} \tau_z + \delta \Delta_{\bm k,\sigma} \tau_x .   
    \label{eq:self_expand}
\end{eqnarray} 
The term proportional to $\tau_y$ is absent in this general expression due to the fact  that the superconducting order parameter is chosen to be real. Substituting Eq.~(\ref{eq:self_expand}) into Eq.~(\ref{eq:gorkov_spin}) we immediately obtain the solution for the Green's function in the form
\begin{equation}
    \begin{gathered}
\hat{{G}}_{\bm k,\sigma} (\omega) = \frac{i \widetilde{\omega}_{\bm k, \sigma} +\widetilde{\xi}_{\bm k, \sigma} \tau_z + \widetilde{\Delta}_{\bm k, \sigma} \tau_x}{(i \widetilde{\omega}_{\bm k, \sigma})^2 - (\widetilde{\xi}_{\bm k, \sigma})^2 - (\widetilde{\Delta}_{\bm k, \sigma})^2} ,
    \end{gathered}
    \label{eq:gorkov_sol_app}
\end{equation}
where
\begin{equation}
    \begin{gathered}
\widetilde{\Delta}_{\bm k, \sigma} (\omega) = \Delta + \delta \Delta_{\bm k,\sigma}(\omega), 
    \end{gathered}
    \label{eq:delta_corr_app}
\end{equation} 
\begin{equation}
    \begin{gathered}
\widetilde{\xi}_{\bm k, \sigma} (\omega) = \xi_{\bm k} + \delta \xi_{\bm k,\sigma}(\omega),
    \end{gathered}
    \label{eq:xi_corr_app}
\end{equation}
\begin{equation}
    \begin{gathered}
i\widetilde {\omega}_{\bm k, \sigma} (\omega) = i \omega - \widetilde{U} \sigma + i\delta \omega_{\bm k,\sigma}(\omega).
    \end{gathered}
    \label{eq:energy_corr_app}
\end{equation}
Substituting Eq.~(\ref{eq:gorkov_sol_app}) into the magnon self-energy Eq.~(\ref{eq:magnon_se_spin}) and taking into account expansion (\ref{eq:self_expand}), we obtain
\begin{eqnarray}
    \delta \Delta_{\bm k,\sigma}(\omega) = - V^2 T \times \nonumber \\
\sum\limits_{\bm q, \Omega} B_{\bm q}(\Omega) \frac{\widetilde {\Delta}_{\bm k,\bar \sigma}}{(i \widetilde \omega_{\bm k,\bar \sigma} - i\sigma \Omega )^2 - \widetilde {\xi}^2_{\bm k-\sigma \bm q,\bar \sigma} - \widetilde {\Delta}_{\bm k, \bar \sigma}^2} ,
    \label{eq:delta_corr_app2}
\end{eqnarray} 
\begin{eqnarray}
    \delta \xi_{\bm k,\sigma}(\omega)= - V^2 T \times \nonumber \\
\sum\limits_{\bm q, \Omega} B_{\bm q}(\Omega) \frac{\widetilde {\xi}^2_{\bm k-\sigma \bm q,\bar \sigma}}{(i \widetilde \omega_{\bm k,\bar \sigma} - i\sigma \Omega )^2 - \widetilde {\xi}^2_{\bm k-\sigma \bm q,\bar \sigma} - \widetilde {\Delta}_{\bm k, \bar \sigma}^2} ,
   \label{eq:xi_corr_app2}
\end{eqnarray}
\begin{eqnarray}
    i\delta \omega_{\bm k,\sigma}(\omega)= V^2 T \times \nonumber \\ \sum\limits_{\bm q, \Omega} B_{\bm q}(\Omega) \frac{i \widetilde \omega_{\bm k,\bar \sigma} - i\sigma \Omega }{(i \widetilde \omega_{\bm k,\bar \sigma} - i\sigma \Omega )^2 - \widetilde {\xi}^2_{\bm k-\sigma \bm q,\bar \sigma} - \widetilde {\Delta}_{\bm k, \bar \sigma}^2} .
\label{eq:energy_corr_app2}
\end{eqnarray}
In principle, Eqs.~(\ref{eq:delta_corr_app2})-(\ref{eq:energy_corr_app2}) represent a system of self-consistency equations for calculation of the magnonic corrections $\delta \Delta_{\bm k, \sigma}$, $\delta \xi_{\bm k, \sigma}$, and $\delta \omega_{\bm k, \sigma}$. However, the smallness of the electron-magnon coupling constant allows us to neglect the corrections on the right-hand sides of these equations. It is equivalent to the non self-consistent Born approximation and leads to Eqs.~(\ref{eq:delta_corr})-(\ref{eq:energy_corr}).

\subsection{Reduction of the Zeeman splitting of the DOS by thermal magnons}

\begin{figure}[tb]
\begin{center}
\includegraphics[width=65mm]{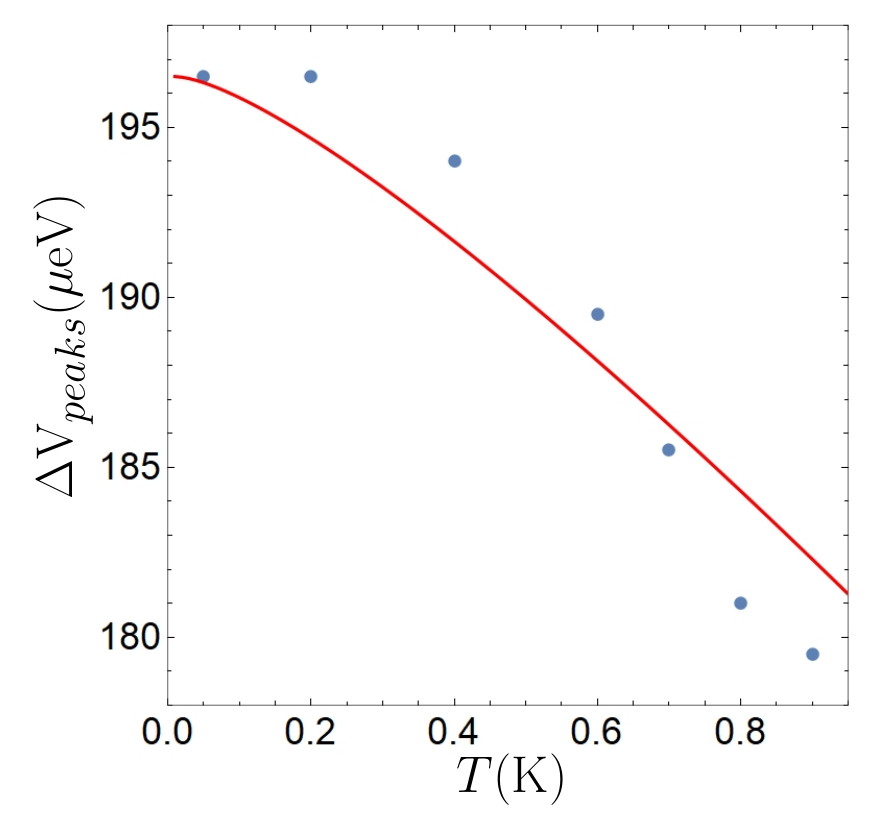}
\caption{Calculated dependence of the Zeeman splitting of the DOS coherence peaks $\Delta V_{peaks}(T)$ accounting for the reduction of the average magnetization of the ferromagnet by thermal magnons (red) and experimental data for $\Delta V_{peaks}(T)$ taken from \cite{Strambini2017} (blue points). $\omega_0 = 0.03$K, $2|\gamma|/M_s = 3.3*10^{-28}m^3$, $D_{EuS} = 3*10^{-42}J*m^2$, $d_{EuS}=5nm$.}
\label{fig:fit_zeeman}
\end{center}
\end{figure}

The suppression of the average magnetization of the ferromagnet by thermal magnons results in the suppression of the spin-splitting field $U$ in the superconductor because it is directly proportional to the average magnetization of the ferromagnet. The renormalized value of the effective spin-splitting field can be calculated as $|\tilde U| = J (M_s-N_m |\gamma|)/(2|\gamma|d_S )$ 
with the magnon density $N_m = (1/S d_{FI})\sum_{\bm q}\left\{\exp[-(\omega_0+Dq^2)/T]-1\right\}^{-1}$ \cite{Getzlaff}. Even without taking into account magnon-assisted electron spin flip processes this reduction of the spin-splitting field should contribute to the temperature dependence of the experimentally observed Zeeman splitting of the DOS coherence peaks. 

In Fig.~\ref{fig:fit_zeeman} we demonstrate the Zeeman splitting of the DOS coherence peaks $\Delta V_{peaks}(T) = 2|\tilde U|$, calculated in the framework of this model, and compare it to the experimental data from \cite{Strambini2017}. The only free parameter here is the value of the gap in the magnonic spectra $\omega_0$, which depends strongly on the magnetic anisotropy, realized in the particular sample. The other parameters are extracted from experimental data. We can see that the dependence $\Delta V_{peaks}(T) $ plotted for $\omega_0 \approx 0.03 K$ is in reasonable agreement with the experimental data. The remaining discrepancy between the calculated dependence $\Delta V_{peaks}(T) $ and the data can be due to other reasons, which are also present in real experiments. For example, weakly bound spins at the interface should also contribute to $\Delta V_{peaks}(T) $, as it was indicated in \cite{Strambini2017}.

\bibliography{magnon_DOS}

\begin{thebibliography}{68}%
\makeatletter
\providecommand \@ifxundefined [1]{%
 \@ifx{#1\undefined}
}%
\providecommand \@ifnum [1]{%
 \ifnum #1\expandafter \@firstoftwo
 \else \expandafter \@secondoftwo
 \fi
}%
\providecommand \@ifx [1]{%
 \ifx #1\expandafter \@firstoftwo
 \else \expandafter \@secondoftwo
 \fi
}%
\providecommand \natexlab [1]{#1}%
\providecommand \enquote  [1]{``#1''}%
\providecommand \bibnamefont  [1]{#1}%
\providecommand \bibfnamefont [1]{#1}%
\providecommand \citenamefont [1]{#1}%
\providecommand \href@noop [0]{\@secondoftwo}%
\providecommand \href [0]{\begingroup \@sanitize@url \@href}%
\providecommand \@href[1]{\@@startlink{#1}\@@href}%
\providecommand \@@href[1]{\endgroup#1\@@endlink}%
\providecommand \@sanitize@url [0]{\catcode `\\12\catcode `\$12\catcode
  `\&12\catcode `\#12\catcode `\^12\catcode `\_12\catcode `\%12\relax}%
\providecommand \@@startlink[1]{}%
\providecommand \@@endlink[0]{}%
\providecommand \url  [0]{\begingroup\@sanitize@url \@url }%
\providecommand \@url [1]{\endgroup\@href {#1}{\urlprefix }}%
\providecommand \urlprefix  [0]{URL }%
\providecommand \Eprint [0]{\href }%
\providecommand \doibase [0]{http://dx.doi.org/}%
\providecommand \selectlanguage [0]{\@gobble}%
\providecommand \bibinfo  [0]{\@secondoftwo}%
\providecommand \bibfield  [0]{\@secondoftwo}%
\providecommand \translation [1]{[#1]}%
\providecommand \BibitemOpen [0]{}%
\providecommand \bibitemStop [0]{}%
\providecommand \bibitemNoStop [0]{.\EOS\space}%
\providecommand \EOS [0]{\spacefactor3000\relax}%
\providecommand \BibitemShut  [1]{\csname bibitem#1\endcsname}%
\let\auto@bib@innerbib\@empty
\bibitem [{\citenamefont {Hao}\ \emph {et~al.}(1990)\citenamefont {Hao},
  \citenamefont {Moodera},\ and\ \citenamefont {Meservey}}]{Hao1990}%
  \BibitemOpen
  \bibfield  {author} {\bibinfo {author} {\bibfnamefont {X.}~\bibnamefont
  {Hao}}, \bibinfo {author} {\bibfnamefont {J.~S.}\ \bibnamefont {Moodera}}, \
  and\ \bibinfo {author} {\bibfnamefont {R.}~\bibnamefont {Meservey}},\ }\href
  {\doibase 10.1103/PhysRevB.42.8235} {\bibfield  {journal} {\bibinfo
  {journal} {Phys. Rev. B}\ }\textbf {\bibinfo {volume} {42}},\ \bibinfo
  {pages} {8235} (\bibinfo {year} {1990})}\BibitemShut {NoStop}%
\bibitem [{\citenamefont {Moodera}\ \emph {et~al.}(1988)\citenamefont
  {Moodera}, \citenamefont {Hao}, \citenamefont {Gibson},\ and\ \citenamefont
  {Meservey}}]{Moodera1988}%
  \BibitemOpen
  \bibfield  {author} {\bibinfo {author} {\bibfnamefont {J.~S.}\ \bibnamefont
  {Moodera}}, \bibinfo {author} {\bibfnamefont {X.}~\bibnamefont {Hao}},
  \bibinfo {author} {\bibfnamefont {G.~A.}\ \bibnamefont {Gibson}}, \ and\
  \bibinfo {author} {\bibfnamefont {R.}~\bibnamefont {Meservey}},\ }\href
  {\doibase 10.1103/PhysRevLett.61.637} {\bibfield  {journal} {\bibinfo
  {journal} {Phys. Rev. Lett.}\ }\textbf {\bibinfo {volume} {61}},\ \bibinfo
  {pages} {637} (\bibinfo {year} {1988})}\BibitemShut {NoStop}%
\bibitem [{\citenamefont {Tedrow}\ \emph {et~al.}(1986)\citenamefont {Tedrow},
  \citenamefont {Tkaczyk},\ and\ \citenamefont {Kumar}}]{Tedrow1986}%
  \BibitemOpen
  \bibfield  {author} {\bibinfo {author} {\bibfnamefont {P.~M.}\ \bibnamefont
  {Tedrow}}, \bibinfo {author} {\bibfnamefont {J.~E.}\ \bibnamefont {Tkaczyk}},
  \ and\ \bibinfo {author} {\bibfnamefont {A.}~\bibnamefont {Kumar}},\ }\href
  {\doibase 10.1103/PhysRevLett.56.1746} {\bibfield  {journal} {\bibinfo
  {journal} {Phys. Rev. Lett.}\ }\textbf {\bibinfo {volume} {56}},\ \bibinfo
  {pages} {1746} (\bibinfo {year} {1986})}\BibitemShut {NoStop}%
\bibitem [{\citenamefont {Meservey}\ and\ \citenamefont
  {Tedrow}(1994)}]{Meservey1994}%
  \BibitemOpen
  \bibfield  {author} {\bibinfo {author} {\bibfnamefont {R.}~\bibnamefont
  {Meservey}}\ and\ \bibinfo {author} {\bibfnamefont {P.~M.}\ \bibnamefont
  {Tedrow}},\ }\href
  {https://www.sciencedirect.com/science/article/pii/0370157394901058}
  {\bibfield  {journal} {\bibinfo  {journal} {Physics Reports}\ }\textbf
  {\bibinfo {volume} {238}},\ \bibinfo {pages} {173} (\bibinfo {year}
  {1994})}\BibitemShut {NoStop}%
\bibitem [{\citenamefont {Linder}\ and\ \citenamefont
  {Robinson}(2015)}]{Linder2015}%
  \BibitemOpen
  \bibfield  {author} {\bibinfo {author} {\bibfnamefont {J.}~\bibnamefont
  {Linder}}\ and\ \bibinfo {author} {\bibfnamefont {J.~W.~A.}\ \bibnamefont
  {Robinson}},\ }\href {\doibase 10.1038/nphys3242} {\bibfield  {journal}
  {\bibinfo  {journal} {Nature Physics}\ }\textbf {\bibinfo {volume} {11}},\
  \bibinfo {pages} {307} (\bibinfo {year} {2015})}\BibitemShut {NoStop}%
\bibitem [{\citenamefont {Eschrig}(2015)}]{Eschrig2015}%
  \BibitemOpen
  \bibfield  {author} {\bibinfo {author} {\bibfnamefont {M.}~\bibnamefont
  {Eschrig}},\ }\href {\doibase 10.1088/0034-4885/78/10/104501} {\bibfield
  {journal} {\bibinfo  {journal} {Reports on Progress in Physics}\ }\textbf
  {\bibinfo {volume} {78}},\ \bibinfo {pages} {104501} (\bibinfo {year}
  {2015})}\BibitemShut {NoStop}%
\bibitem [{\citenamefont {Bergeret}\ \emph {et~al.}(2018)\citenamefont
  {Bergeret}, \citenamefont {Silaev}, \citenamefont {Virtanen},\ and\
  \citenamefont {Heikkil\"a}}]{Bergeret2018review}%
  \BibitemOpen
  \bibfield  {author} {\bibinfo {author} {\bibfnamefont {F.~S.}\ \bibnamefont
  {Bergeret}}, \bibinfo {author} {\bibfnamefont {M.}~\bibnamefont {Silaev}},
  \bibinfo {author} {\bibfnamefont {P.}~\bibnamefont {Virtanen}}, \ and\
  \bibinfo {author} {\bibfnamefont {T.~T.}\ \bibnamefont {Heikkil\"a}},\ }\href
  {\doibase 10.1103/RevModPhys.90.041001} {\bibfield  {journal} {\bibinfo
  {journal} {Rev. Mod. Phys.}\ }\textbf {\bibinfo {volume} {90}},\ \bibinfo
  {pages} {041001} (\bibinfo {year} {2018})}\BibitemShut {NoStop}%
\bibitem [{\citenamefont {Heikkil{\"a}}\ \emph {et~al.}(2019)\citenamefont
  {Heikkil{\"a}}, \citenamefont {Silaev}, \citenamefont {Virtanen},\ and\
  \citenamefont {Bergeret}}]{Heikkila2019review}%
  \BibitemOpen
  \bibfield  {author} {\bibinfo {author} {\bibfnamefont {T.~T.}\ \bibnamefont
  {Heikkil{\"a}}}, \bibinfo {author} {\bibfnamefont {M.}~\bibnamefont
  {Silaev}}, \bibinfo {author} {\bibfnamefont {P.}~\bibnamefont {Virtanen}}, \
  and\ \bibinfo {author} {\bibfnamefont {F.~S.}\ \bibnamefont {Bergeret}},\
  }\href {https://www.sciencedirect.com/science/article/pii/S0079681619300115}
  {\bibfield  {journal} {\bibinfo  {journal} {Progress in Surface Science}\
  }\textbf {\bibinfo {volume} {94}},\ \bibinfo {pages} {100540} (\bibinfo
  {year} {2019})}\BibitemShut {NoStop}%
\bibitem [{\citenamefont {Machon}\ \emph {et~al.}(2013)\citenamefont {Machon},
  \citenamefont {Eschrig},\ and\ \citenamefont {Belzig}}]{Machon2013}%
  \BibitemOpen
  \bibfield  {author} {\bibinfo {author} {\bibfnamefont {P.}~\bibnamefont
  {Machon}}, \bibinfo {author} {\bibfnamefont {M.}~\bibnamefont {Eschrig}}, \
  and\ \bibinfo {author} {\bibfnamefont {W.}~\bibnamefont {Belzig}},\ }\href
  {\doibase 10.1103/PhysRevLett.110.047002} {\bibfield  {journal} {\bibinfo
  {journal} {Phys. Rev. Lett.}\ }\textbf {\bibinfo {volume} {110}},\ \bibinfo
  {pages} {047002} (\bibinfo {year} {2013})}\BibitemShut {NoStop}%
\bibitem [{\citenamefont {Ozaeta}\ \emph {et~al.}(2014)\citenamefont {Ozaeta},
  \citenamefont {Virtanen}, \citenamefont {Bergeret},\ and\ \citenamefont
  {Heikkil\"a}}]{Ozaeta2014}%
  \BibitemOpen
  \bibfield  {author} {\bibinfo {author} {\bibfnamefont {A.}~\bibnamefont
  {Ozaeta}}, \bibinfo {author} {\bibfnamefont {P.}~\bibnamefont {Virtanen}},
  \bibinfo {author} {\bibfnamefont {F.~S.}\ \bibnamefont {Bergeret}}, \ and\
  \bibinfo {author} {\bibfnamefont {T.~T.}\ \bibnamefont {Heikkil\"a}},\ }\href
  {\doibase 10.1103/PhysRevLett.112.057001} {\bibfield  {journal} {\bibinfo
  {journal} {Phys. Rev. Lett.}\ }\textbf {\bibinfo {volume} {112}},\ \bibinfo
  {pages} {057001} (\bibinfo {year} {2014})}\BibitemShut {NoStop}%
\bibitem [{\citenamefont {Kolenda}\ \emph
  {et~al.}(2016{\natexlab{a}})\citenamefont {Kolenda}, \citenamefont {Wolf},\
  and\ \citenamefont {Beckmann}}]{Kolenda2016}%
  \BibitemOpen
  \bibfield  {author} {\bibinfo {author} {\bibfnamefont {S.}~\bibnamefont
  {Kolenda}}, \bibinfo {author} {\bibfnamefont {M.~J.}\ \bibnamefont {Wolf}}, \
  and\ \bibinfo {author} {\bibfnamefont {D.}~\bibnamefont {Beckmann}},\ }\href
  {\doibase 10.1103/PhysRevLett.116.097001} {\bibfield  {journal} {\bibinfo
  {journal} {Phys. Rev. Lett.}\ }\textbf {\bibinfo {volume} {116}},\ \bibinfo
  {pages} {097001} (\bibinfo {year} {2016}{\natexlab{a}})}\BibitemShut
  {NoStop}%
\bibitem [{\citenamefont {Kolenda}\ \emph
  {et~al.}(2016{\natexlab{b}})\citenamefont {Kolenda}, \citenamefont {Machon},
  \citenamefont {Beckmann},\ and\ \citenamefont {Belzig}}]{Kolenda2016_2}%
  \BibitemOpen
  \bibfield  {author} {\bibinfo {author} {\bibfnamefont {S.}~\bibnamefont
  {Kolenda}}, \bibinfo {author} {\bibfnamefont {P.}~\bibnamefont {Machon}},
  \bibinfo {author} {\bibfnamefont {D.}~\bibnamefont {Beckmann}}, \ and\
  \bibinfo {author} {\bibfnamefont {W.}~\bibnamefont {Belzig}},\ }\href
  {\doibase 10.3762/bjnano.7.152} {\bibfield  {journal} {\bibinfo  {journal}
  {Beilstein J. Nanotechnol.}\ }\textbf {\bibinfo {volume} {7}},\ \bibinfo
  {pages} {1579} (\bibinfo {year} {2016}{\natexlab{b}})}\BibitemShut {NoStop}%
\bibitem [{\citenamefont {Giazotto}\ \emph {et~al.}(2014)\citenamefont
  {Giazotto}, \citenamefont {Robinson}, \citenamefont {Moodera},\ and\
  \citenamefont {Bergeret}}]{Giazotto2014}%
  \BibitemOpen
  \bibfield  {author} {\bibinfo {author} {\bibfnamefont {F.}~\bibnamefont
  {Giazotto}}, \bibinfo {author} {\bibfnamefont {J.~W.~A.}\ \bibnamefont
  {Robinson}}, \bibinfo {author} {\bibfnamefont {J.~S.}\ \bibnamefont
  {Moodera}}, \ and\ \bibinfo {author} {\bibfnamefont {F.~S.}\ \bibnamefont
  {Bergeret}},\ }\href {\doibase 10.1063/1.4893443} {\bibfield  {journal}
  {\bibinfo  {journal} {Applied Physics Letters}\ }\textbf {\bibinfo {volume}
  {105}},\ \bibinfo {pages} {062602} (\bibinfo {year} {2014})}\BibitemShut
  {NoStop}%
\bibitem [{\citenamefont {Giazotto}\ \emph
  {et~al.}(2015{\natexlab{a}})\citenamefont {Giazotto}, \citenamefont
  {Heikkil\"a},\ and\ \citenamefont {Bergeret}}]{Giazotto2015}%
  \BibitemOpen
  \bibfield  {author} {\bibinfo {author} {\bibfnamefont {F.}~\bibnamefont
  {Giazotto}}, \bibinfo {author} {\bibfnamefont {T.~T.}\ \bibnamefont
  {Heikkil\"a}}, \ and\ \bibinfo {author} {\bibfnamefont {F.~S.}\ \bibnamefont
  {Bergeret}},\ }\href {\doibase 10.1103/PhysRevLett.114.067001} {\bibfield
  {journal} {\bibinfo  {journal} {Phys. Rev. Lett.}\ }\textbf {\bibinfo
  {volume} {114}},\ \bibinfo {pages} {067001} (\bibinfo {year}
  {2015}{\natexlab{a}})}\BibitemShut {NoStop}%
\bibitem [{\citenamefont {Machon}\ \emph {et~al.}(2014)\citenamefont {Machon},
  \citenamefont {Eschrig},\ and\ \citenamefont {Belzig}}]{Machon2014}%
  \BibitemOpen
  \bibfield  {author} {\bibinfo {author} {\bibfnamefont {P.}~\bibnamefont
  {Machon}}, \bibinfo {author} {\bibfnamefont {M.}~\bibnamefont {Eschrig}}, \
  and\ \bibinfo {author} {\bibfnamefont {W.}~\bibnamefont {Belzig}},\ }\href
  {\doibase 10.1088/1367-2630/16/7/073002} {\bibfield  {journal} {\bibinfo
  {journal} {New Journal of Physics}\ }\textbf {\bibinfo {volume} {16}},\
  \bibinfo {pages} {073002} (\bibinfo {year} {2014})}\BibitemShut {NoStop}%
\bibitem [{\citenamefont {Kolenda}\ \emph {et~al.}(2017)\citenamefont
  {Kolenda}, \citenamefont {S\"urgers}, \citenamefont {Fischer},\ and\
  \citenamefont {Beckmann}}]{Kolenda2017}%
  \BibitemOpen
  \bibfield  {author} {\bibinfo {author} {\bibfnamefont {S.}~\bibnamefont
  {Kolenda}}, \bibinfo {author} {\bibfnamefont {C.}~\bibnamefont {S\"urgers}},
  \bibinfo {author} {\bibfnamefont {G.}~\bibnamefont {Fischer}}, \ and\
  \bibinfo {author} {\bibfnamefont {D.}~\bibnamefont {Beckmann}},\ }\href
  {\doibase 10.1103/PhysRevB.95.224505} {\bibfield  {journal} {\bibinfo
  {journal} {Phys. Rev. B}\ }\textbf {\bibinfo {volume} {95}},\ \bibinfo
  {pages} {224505} (\bibinfo {year} {2017})}\BibitemShut {NoStop}%
\bibitem [{\citenamefont {Rezaei}\ \emph {et~al.}(2018)\citenamefont {Rezaei},
  \citenamefont {Kamra}, \citenamefont {Machon},\ and\ \citenamefont
  {Belzig}}]{Rezaei2018}%
  \BibitemOpen
  \bibfield  {author} {\bibinfo {author} {\bibfnamefont {A.}~\bibnamefont
  {Rezaei}}, \bibinfo {author} {\bibfnamefont {A.}~\bibnamefont {Kamra}},
  \bibinfo {author} {\bibfnamefont {P.}~\bibnamefont {Machon}}, \ and\ \bibinfo
  {author} {\bibfnamefont {W.}~\bibnamefont {Belzig}},\ }\href {\doibase
  10.1088/1367-2630/aad2a3} {\bibfield  {journal} {\bibinfo  {journal} {New
  Journal of Physics}\ }\textbf {\bibinfo {volume} {20}},\ \bibinfo {pages}
  {073034} (\bibinfo {year} {2018})}\BibitemShut {NoStop}%
\bibitem [{\citenamefont {Linder}\ and\ \citenamefont
  {Bathen}(2016)}]{Linder2016}%
  \BibitemOpen
  \bibfield  {author} {\bibinfo {author} {\bibfnamefont {J.}~\bibnamefont
  {Linder}}\ and\ \bibinfo {author} {\bibfnamefont {M.~E.}\ \bibnamefont
  {Bathen}},\ }\href {\doibase 10.1103/PhysRevB.93.224509} {\bibfield
  {journal} {\bibinfo  {journal} {Phys. Rev. B}\ }\textbf {\bibinfo {volume}
  {93}},\ \bibinfo {pages} {224509} (\bibinfo {year} {2016})}\BibitemShut
  {NoStop}%
\bibitem [{\citenamefont {Bobkova}\ and\ \citenamefont
  {Bobkov}(2017)}]{Bobkova2017}%
  \BibitemOpen
  \bibfield  {author} {\bibinfo {author} {\bibfnamefont {I.~V.}\ \bibnamefont
  {Bobkova}}\ and\ \bibinfo {author} {\bibfnamefont {A.~M.}\ \bibnamefont
  {Bobkov}},\ }\href {\doibase 10.1103/PhysRevB.96.104515} {\bibfield
  {journal} {\bibinfo  {journal} {Phys. Rev. B}\ }\textbf {\bibinfo {volume}
  {96}},\ \bibinfo {pages} {104515} (\bibinfo {year} {2017})}\BibitemShut
  {NoStop}%
\bibitem [{\citenamefont {Bobkova}\ \emph {et~al.}(2021)\citenamefont
  {Bobkova}, \citenamefont {Bobkov},\ and\ \citenamefont
  {Belzig}}]{Bobkova2021}%
  \BibitemOpen
  \bibfield  {author} {\bibinfo {author} {\bibfnamefont {I.~V.}\ \bibnamefont
  {Bobkova}}, \bibinfo {author} {\bibfnamefont {A.~M.}\ \bibnamefont {Bobkov}},
  \ and\ \bibinfo {author} {\bibfnamefont {W.}~\bibnamefont {Belzig}},\ }\href
  {\doibase 10.1103/PhysRevB.103.L020503} {\bibfield  {journal} {\bibinfo
  {journal} {Phys. Rev. B}\ }\textbf {\bibinfo {volume} {103}},\ \bibinfo
  {pages} {L020503} (\bibinfo {year} {2021})}\BibitemShut {NoStop}%
\bibitem [{\citenamefont {Huertas-Hernando}\ \emph {et~al.}(2002)\citenamefont
  {Huertas-Hernando}, \citenamefont {Nazarov},\ and\ \citenamefont
  {Belzig}}]{Huertas-Hernando2002}%
  \BibitemOpen
  \bibfield  {author} {\bibinfo {author} {\bibfnamefont {D.}~\bibnamefont
  {Huertas-Hernando}}, \bibinfo {author} {\bibfnamefont {Y.~V.}\ \bibnamefont
  {Nazarov}}, \ and\ \bibinfo {author} {\bibfnamefont {W.}~\bibnamefont
  {Belzig}},\ }\href {\doibase 10.1103/PhysRevLett.88.047003} {\bibfield
  {journal} {\bibinfo  {journal} {Phys. Rev. Lett.}\ }\textbf {\bibinfo
  {volume} {88}},\ \bibinfo {pages} {047003} (\bibinfo {year}
  {2002})}\BibitemShut {NoStop}%
\bibitem [{\citenamefont {Giazotto}\ \emph
  {et~al.}(2006{\natexlab{a}})\citenamefont {Giazotto}, \citenamefont {Taddei},
  \citenamefont {Fazio},\ and\ \citenamefont {Beltram}}]{Giazotto2006}%
  \BibitemOpen
  \bibfield  {author} {\bibinfo {author} {\bibfnamefont {F.}~\bibnamefont
  {Giazotto}}, \bibinfo {author} {\bibfnamefont {F.}~\bibnamefont {Taddei}},
  \bibinfo {author} {\bibfnamefont {R.}~\bibnamefont {Fazio}}, \ and\ \bibinfo
  {author} {\bibfnamefont {F.}~\bibnamefont {Beltram}},\ }\href {\doibase
  10.1063/1.2220001} {\bibfield  {journal} {\bibinfo  {journal} {Applied
  Physics Letters}\ }\textbf {\bibinfo {volume} {89}},\ \bibinfo {pages}
  {022505} (\bibinfo {year} {2006}{\natexlab{a}})}\BibitemShut {NoStop}%
\bibitem [{\citenamefont {Giazotto}\ and\ \citenamefont
  {Taddei}(2008)}]{Giazotto2008}%
  \BibitemOpen
  \bibfield  {author} {\bibinfo {author} {\bibfnamefont {F.}~\bibnamefont
  {Giazotto}}\ and\ \bibinfo {author} {\bibfnamefont {F.}~\bibnamefont
  {Taddei}},\ }\href {\doibase 10.1103/PhysRevB.77.132501} {\bibfield
  {journal} {\bibinfo  {journal} {Phys. Rev. B}\ }\textbf {\bibinfo {volume}
  {77}},\ \bibinfo {pages} {132501} (\bibinfo {year} {2008})}\BibitemShut
  {NoStop}%
\bibitem [{\citenamefont {Giazotto}\ and\ \citenamefont
  {Bergeret}(2013)}]{Giazotto2013}%
  \BibitemOpen
  \bibfield  {author} {\bibinfo {author} {\bibfnamefont {F.}~\bibnamefont
  {Giazotto}}\ and\ \bibinfo {author} {\bibfnamefont {F.~S.}\ \bibnamefont
  {Bergeret}},\ }\href {\doibase 10.1063/1.4802953} {\bibfield  {journal}
  {\bibinfo  {journal} {Applied Physics Letters}\ }\textbf {\bibinfo {volume}
  {102}},\ \bibinfo {pages} {162406} (\bibinfo {year} {2013})}\BibitemShut
  {NoStop}%
\bibitem [{\citenamefont {Strambini}\ \emph {et~al.}(2017)\citenamefont
  {Strambini}, \citenamefont {Golovach}, \citenamefont {De~Simoni},
  \citenamefont {Moodera}, \citenamefont {Bergeret},\ and\ \citenamefont
  {Giazotto}}]{Strambini2017}%
  \BibitemOpen
  \bibfield  {author} {\bibinfo {author} {\bibfnamefont {E.}~\bibnamefont
  {Strambini}}, \bibinfo {author} {\bibfnamefont {V.~N.}\ \bibnamefont
  {Golovach}}, \bibinfo {author} {\bibfnamefont {G.}~\bibnamefont {De~Simoni}},
  \bibinfo {author} {\bibfnamefont {J.~S.}\ \bibnamefont {Moodera}}, \bibinfo
  {author} {\bibfnamefont {F.~S.}\ \bibnamefont {Bergeret}}, \ and\ \bibinfo
  {author} {\bibfnamefont {F.}~\bibnamefont {Giazotto}},\ }\href {\doibase
  10.1103/PhysRevMaterials.1.054402} {\bibfield  {journal} {\bibinfo  {journal}
  {Phys. Rev. Mater.}\ }\textbf {\bibinfo {volume} {1}},\ \bibinfo {pages}
  {054402} (\bibinfo {year} {2017})}\BibitemShut {NoStop}%
\bibitem [{\citenamefont {Giazotto}\ \emph
  {et~al.}(2006{\natexlab{b}})\citenamefont {Giazotto}, \citenamefont
  {Heikkil\"a}, \citenamefont {Luukanen}, \citenamefont {Savin},\ and\
  \citenamefont {Pekola}}]{Giazotto2006review}%
  \BibitemOpen
  \bibfield  {author} {\bibinfo {author} {\bibfnamefont {F.}~\bibnamefont
  {Giazotto}}, \bibinfo {author} {\bibfnamefont {T.~T.}\ \bibnamefont
  {Heikkil\"a}}, \bibinfo {author} {\bibfnamefont {A.}~\bibnamefont
  {Luukanen}}, \bibinfo {author} {\bibfnamefont {A.~M.}\ \bibnamefont {Savin}},
  \ and\ \bibinfo {author} {\bibfnamefont {J.~P.}\ \bibnamefont {Pekola}},\
  }\href {\doibase 10.1103/RevModPhys.78.217} {\bibfield  {journal} {\bibinfo
  {journal} {Rev. Mod. Phys.}\ }\textbf {\bibinfo {volume} {78}},\ \bibinfo
  {pages} {217} (\bibinfo {year} {2006}{\natexlab{b}})}\BibitemShut {NoStop}%
\bibitem [{\citenamefont {Kawabata}\ \emph {et~al.}(2013)\citenamefont
  {Kawabata}, \citenamefont {Ozaeta}, \citenamefont {Vasenko}, \citenamefont
  {Hekking},\ and\ \citenamefont {Sebasti{\'a}n~Bergeret}}]{Kawabata2013}%
  \BibitemOpen
  \bibfield  {author} {\bibinfo {author} {\bibfnamefont {S.}~\bibnamefont
  {Kawabata}}, \bibinfo {author} {\bibfnamefont {A.}~\bibnamefont {Ozaeta}},
  \bibinfo {author} {\bibfnamefont {A.~S.}\ \bibnamefont {Vasenko}}, \bibinfo
  {author} {\bibfnamefont {F.~W.~J.}\ \bibnamefont {Hekking}}, \ and\ \bibinfo
  {author} {\bibfnamefont {F.}~\bibnamefont {Sebasti{\'a}n~Bergeret}},\ }\href
  {\doibase 10.1063/1.4813599} {\bibfield  {journal} {\bibinfo  {journal}
  {Applied Physics Letters}\ }\textbf {\bibinfo {volume} {103}},\ \bibinfo
  {pages} {032602} (\bibinfo {year} {2013})}\BibitemShut {NoStop}%
\bibitem [{\citenamefont {Giazotto}\ \emph
  {et~al.}(2015{\natexlab{b}})\citenamefont {Giazotto}, \citenamefont
  {Solinas}, \citenamefont {Braggio},\ and\ \citenamefont
  {Bergeret}}]{Giazotto2015_2}%
  \BibitemOpen
  \bibfield  {author} {\bibinfo {author} {\bibfnamefont {F.}~\bibnamefont
  {Giazotto}}, \bibinfo {author} {\bibfnamefont {P.}~\bibnamefont {Solinas}},
  \bibinfo {author} {\bibfnamefont {A.}~\bibnamefont {Braggio}}, \ and\
  \bibinfo {author} {\bibfnamefont {F.~S.}\ \bibnamefont {Bergeret}},\ }\href
  {\doibase 10.1103/PhysRevApplied.4.044016} {\bibfield  {journal} {\bibinfo
  {journal} {Phys. Rev. Appl.}\ }\textbf {\bibinfo {volume} {4}},\ \bibinfo
  {pages} {044016} (\bibinfo {year} {2015}{\natexlab{b}})}\BibitemShut
  {NoStop}%
\bibitem [{\citenamefont {Heikkil\"a}\ \emph {et~al.}(2018)\citenamefont
  {Heikkil\"a}, \citenamefont {Ojaj\"arvi}, \citenamefont {Maasilta},
  \citenamefont {Strambini}, \citenamefont {Giazotto},\ and\ \citenamefont
  {Bergeret}}]{Heikkila2018}%
  \BibitemOpen
  \bibfield  {author} {\bibinfo {author} {\bibfnamefont {T.~T.}\ \bibnamefont
  {Heikkil\"a}}, \bibinfo {author} {\bibfnamefont {R.}~\bibnamefont
  {Ojaj\"arvi}}, \bibinfo {author} {\bibfnamefont {I.~J.}\ \bibnamefont
  {Maasilta}}, \bibinfo {author} {\bibfnamefont {E.}~\bibnamefont {Strambini}},
  \bibinfo {author} {\bibfnamefont {F.}~\bibnamefont {Giazotto}}, \ and\
  \bibinfo {author} {\bibfnamefont {F.~S.}\ \bibnamefont {Bergeret}},\ }\href
  {\doibase 10.1103/PhysRevApplied.10.034053} {\bibfield  {journal} {\bibinfo
  {journal} {Phys. Rev. Appl.}\ }\textbf {\bibinfo {volume} {10}},\ \bibinfo
  {pages} {034053} (\bibinfo {year} {2018})}\BibitemShut {NoStop}%
\bibitem [{\citenamefont {Geng}\ \emph {et~al.}(2023)\citenamefont {Geng},
  \citenamefont {Hijano}, \citenamefont {Ilic}, \citenamefont {Ilyn},
  \citenamefont {Maasilta}, \citenamefont {Monfardini}, \citenamefont {Spies},
  \citenamefont {Strambini}, \citenamefont {Virtanen}, \citenamefont {Calvo},
  \citenamefont {Gonzalez-Orellana}, \citenamefont {Helenius}, \citenamefont
  {Khorshidian}, \citenamefont {Levartoski~de Araujo}, \citenamefont
  {Levy-Bertrand}, \citenamefont {Rogero}, \citenamefont {Giazotto},
  \citenamefont {Bergeret},\ and\ \citenamefont {Heikkil\"a}}]{Geng2023}%
  \BibitemOpen
  \bibfield  {author} {\bibinfo {author} {\bibfnamefont {Z.}~\bibnamefont
  {Geng}}, \bibinfo {author} {\bibfnamefont {A.}~\bibnamefont {Hijano}},
  \bibinfo {author} {\bibfnamefont {S.}~\bibnamefont {Ilic}}, \bibinfo {author}
  {\bibfnamefont {M.}~\bibnamefont {Ilyn}}, \bibinfo {author} {\bibfnamefont
  {I.~J.}\ \bibnamefont {Maasilta}}, \bibinfo {author} {\bibfnamefont
  {A.}~\bibnamefont {Monfardini}}, \bibinfo {author} {\bibfnamefont
  {M.}~\bibnamefont {Spies}}, \bibinfo {author} {\bibfnamefont
  {E.}~\bibnamefont {Strambini}}, \bibinfo {author} {\bibfnamefont
  {P.}~\bibnamefont {Virtanen}}, \bibinfo {author} {\bibfnamefont
  {M.}~\bibnamefont {Calvo}}, \bibinfo {author} {\bibfnamefont
  {C.}~\bibnamefont {Gonzalez-Orellana}}, \bibinfo {author} {\bibfnamefont
  {A.~P.}\ \bibnamefont {Helenius}}, \bibinfo {author} {\bibfnamefont
  {S.}~\bibnamefont {Khorshidian}}, \bibinfo {author} {\bibfnamefont {C.~I.}\
  \bibnamefont {Levartoski~de Araujo}}, \bibinfo {author} {\bibfnamefont
  {F.}~\bibnamefont {Levy-Bertrand}}, \bibinfo {author} {\bibfnamefont
  {C.}~\bibnamefont {Rogero}}, \bibinfo {author} {\bibfnamefont
  {F.}~\bibnamefont {Giazotto}}, \bibinfo {author} {\bibfnamefont {F.~S.}\
  \bibnamefont {Bergeret}}, \ and\ \bibinfo {author} {\bibfnamefont {T.~T.}\
  \bibnamefont {Heikkil\"a}},\ }\href@noop {} {\enquote {\bibinfo {title}
  {Superconductor-ferromagnet hybrids for non-reciprocal electronics and
  detectors},}\ } (\bibinfo {year} {2023}),\ \Eprint
  {http://arxiv.org/abs/2302.12732} {arXiv:2302.12732 [cond-mat.supr-con]}
  \BibitemShut {NoStop}%
\bibitem [{\citenamefont {Volkov}\ \emph {et~al.}(2005)\citenamefont {Volkov},
  \citenamefont {Fominov},\ and\ \citenamefont {Efetov}}]{Volkov2005}%
  \BibitemOpen
  \bibfield  {author} {\bibinfo {author} {\bibfnamefont {A.~F.}\ \bibnamefont
  {Volkov}}, \bibinfo {author} {\bibfnamefont {Y.~V.}\ \bibnamefont {Fominov}},
  \ and\ \bibinfo {author} {\bibfnamefont {K.~B.}\ \bibnamefont {Efetov}},\
  }\href {\doibase 10.1103/PhysRevB.72.184504} {\bibfield  {journal} {\bibinfo
  {journal} {Phys. Rev. B}\ }\textbf {\bibinfo {volume} {72}},\ \bibinfo
  {pages} {184504} (\bibinfo {year} {2005})}\BibitemShut {NoStop}%
\bibitem [{\citenamefont {Golubov}\ \emph {et~al.}(2005)\citenamefont
  {Golubov}, \citenamefont {Kupriyanov},\ and\ \citenamefont
  {Siegel}}]{Golubov2005}%
  \BibitemOpen
  \bibfield  {author} {\bibinfo {author} {\bibfnamefont {A.~A.}\ \bibnamefont
  {Golubov}}, \bibinfo {author} {\bibfnamefont {M.~Y.}\ \bibnamefont
  {Kupriyanov}}, \ and\ \bibinfo {author} {\bibfnamefont {M.}~\bibnamefont
  {Siegel}},\ }\href {\doibase 10.1134/1.1914877} {\bibfield  {journal}
  {\bibinfo  {journal} {JETP Letters}\ }\textbf {\bibinfo {volume} {81}},\
  \bibinfo {pages} {180} (\bibinfo {year} {2005})}\BibitemShut {NoStop}%
\bibitem [{\citenamefont {Ivanov}\ and\ \citenamefont
  {Fominov}(2006)}]{Ivanov2007}%
  \BibitemOpen
  \bibfield  {author} {\bibinfo {author} {\bibfnamefont {D.~A.}\ \bibnamefont
  {Ivanov}}\ and\ \bibinfo {author} {\bibfnamefont {Y.~V.}\ \bibnamefont
  {Fominov}},\ }\href {\doibase 10.1103/PhysRevB.73.214524} {\bibfield
  {journal} {\bibinfo  {journal} {Phys. Rev. B}\ }\textbf {\bibinfo {volume}
  {73}},\ \bibinfo {pages} {214524} (\bibinfo {year} {2006})}\BibitemShut
  {NoStop}%
\bibitem [{\citenamefont {Di~Bernardo}\ \emph {et~al.}(2015)\citenamefont
  {Di~Bernardo}, \citenamefont {Diesch}, \citenamefont {Gu}, \citenamefont
  {Linder}, \citenamefont {Divitini}, \citenamefont {Ducati}, \citenamefont
  {Scheer}, \citenamefont {Blamire},\ and\ \citenamefont
  {Robinson}}]{DiBernardo2015}%
  \BibitemOpen
  \bibfield  {author} {\bibinfo {author} {\bibfnamefont {A.}~\bibnamefont
  {Di~Bernardo}}, \bibinfo {author} {\bibfnamefont {S.}~\bibnamefont {Diesch}},
  \bibinfo {author} {\bibfnamefont {Y.}~\bibnamefont {Gu}}, \bibinfo {author}
  {\bibfnamefont {J.}~\bibnamefont {Linder}}, \bibinfo {author} {\bibfnamefont
  {G.}~\bibnamefont {Divitini}}, \bibinfo {author} {\bibfnamefont
  {C.}~\bibnamefont {Ducati}}, \bibinfo {author} {\bibfnamefont
  {E.}~\bibnamefont {Scheer}}, \bibinfo {author} {\bibfnamefont {M.~G.}\
  \bibnamefont {Blamire}}, \ and\ \bibinfo {author} {\bibfnamefont {J.~W.~A.}\
  \bibnamefont {Robinson}},\ }\href {\doibase 10.1038/ncomms9053} {\bibfield
  {journal} {\bibinfo  {journal} {Nature Communications}\ }\textbf {\bibinfo
  {volume} {6}},\ \bibinfo {pages} {8053} (\bibinfo {year} {2015})}\BibitemShut
  {NoStop}%
\bibitem [{\citenamefont {Diesch}\ \emph {et~al.}(2018)\citenamefont {Diesch},
  \citenamefont {Machon}, \citenamefont {Wolz}, \citenamefont {S{\"u}rgers},
  \citenamefont {Beckmann}, \citenamefont {Belzig},\ and\ \citenamefont
  {Scheer}}]{Diesch2018}%
  \BibitemOpen
  \bibfield  {author} {\bibinfo {author} {\bibfnamefont {S.}~\bibnamefont
  {Diesch}}, \bibinfo {author} {\bibfnamefont {P.}~\bibnamefont {Machon}},
  \bibinfo {author} {\bibfnamefont {M.}~\bibnamefont {Wolz}}, \bibinfo {author}
  {\bibfnamefont {C.}~\bibnamefont {S{\"u}rgers}}, \bibinfo {author}
  {\bibfnamefont {D.}~\bibnamefont {Beckmann}}, \bibinfo {author}
  {\bibfnamefont {W.}~\bibnamefont {Belzig}}, \ and\ \bibinfo {author}
  {\bibfnamefont {E.}~\bibnamefont {Scheer}},\ }\href {\doibase
  10.1038/s41467-018-07597-w} {\bibfield  {journal} {\bibinfo  {journal}
  {Nature Communications}\ }\textbf {\bibinfo {volume} {9}},\ \bibinfo {pages}
  {5248} (\bibinfo {year} {2018})}\BibitemShut {NoStop}%
\bibitem [{\citenamefont {Bobkova}\ and\ \citenamefont
  {Bobkov}(2019)}]{Bobkova2019}%
  \BibitemOpen
  \bibfield  {author} {\bibinfo {author} {\bibfnamefont {I.~V.}\ \bibnamefont
  {Bobkova}}\ and\ \bibinfo {author} {\bibfnamefont {A.~M.}\ \bibnamefont
  {Bobkov}},\ }\href {\doibase 10.1134/S0021364019010016} {\bibfield  {journal}
  {\bibinfo  {journal} {JETP Letters}\ }\textbf {\bibinfo {volume} {109}},\
  \bibinfo {pages} {57} (\bibinfo {year} {2019})}\BibitemShut {NoStop}%
\bibitem [{\citenamefont {Bobkova}\ \emph {et~al.}(2019)\citenamefont
  {Bobkova}, \citenamefont {Bobkov},\ and\ \citenamefont
  {Belzig}}]{Bobkova2019_2}%
  \BibitemOpen
  \bibfield  {author} {\bibinfo {author} {\bibfnamefont {I.~V.}\ \bibnamefont
  {Bobkova}}, \bibinfo {author} {\bibfnamefont {A.~M.}\ \bibnamefont {Bobkov}},
  \ and\ \bibinfo {author} {\bibfnamefont {W.}~\bibnamefont {Belzig}},\ }\href
  {\doibase 10.1088/1367-2630/ab0e20} {\bibfield  {journal} {\bibinfo
  {journal} {New Journal of Physics}\ }\textbf {\bibinfo {volume} {21}},\
  \bibinfo {pages} {043001} (\bibinfo {year} {2019})}\BibitemShut {NoStop}%
\bibitem [{\citenamefont {Tserkovnyak}\ \emph {et~al.}(2005)\citenamefont
  {Tserkovnyak}, \citenamefont {Brataas}, \citenamefont {Bauer},\ and\
  \citenamefont {Halperin}}]{Tserkovnyak2005}%
  \BibitemOpen
  \bibfield  {author} {\bibinfo {author} {\bibfnamefont {Y.}~\bibnamefont
  {Tserkovnyak}}, \bibinfo {author} {\bibfnamefont {A.}~\bibnamefont
  {Brataas}}, \bibinfo {author} {\bibfnamefont {G.~E.~W.}\ \bibnamefont
  {Bauer}}, \ and\ \bibinfo {author} {\bibfnamefont {B.~I.}\ \bibnamefont
  {Halperin}},\ }\href {\doibase 10.1103/RevModPhys.77.1375} {\bibfield
  {journal} {\bibinfo  {journal} {Rev. Mod. Phys.}\ }\textbf {\bibinfo {volume}
  {77}},\ \bibinfo {pages} {1375} (\bibinfo {year} {2005})}\BibitemShut
  {NoStop}%
\bibitem [{\citenamefont {Ohnuma}\ \emph {et~al.}(2014)\citenamefont {Ohnuma},
  \citenamefont {Adachi}, \citenamefont {Saitoh},\ and\ \citenamefont
  {Maekawa}}]{Ohnuma2014}%
  \BibitemOpen
  \bibfield  {author} {\bibinfo {author} {\bibfnamefont {Y.}~\bibnamefont
  {Ohnuma}}, \bibinfo {author} {\bibfnamefont {H.}~\bibnamefont {Adachi}},
  \bibinfo {author} {\bibfnamefont {E.}~\bibnamefont {Saitoh}}, \ and\ \bibinfo
  {author} {\bibfnamefont {S.}~\bibnamefont {Maekawa}},\ }\href {\doibase
  10.1103/PhysRevB.89.174417} {\bibfield  {journal} {\bibinfo  {journal} {Phys.
  Rev. B}\ }\textbf {\bibinfo {volume} {89}},\ \bibinfo {pages} {174417}
  (\bibinfo {year} {2014})}\BibitemShut {NoStop}%
\bibitem [{\citenamefont {Bell}\ \emph {et~al.}(2008)\citenamefont {Bell},
  \citenamefont {Milikisyants}, \citenamefont {Huber},\ and\ \citenamefont
  {Aarts}}]{Bell2008}%
  \BibitemOpen
  \bibfield  {author} {\bibinfo {author} {\bibfnamefont {C.}~\bibnamefont
  {Bell}}, \bibinfo {author} {\bibfnamefont {S.}~\bibnamefont {Milikisyants}},
  \bibinfo {author} {\bibfnamefont {M.}~\bibnamefont {Huber}}, \ and\ \bibinfo
  {author} {\bibfnamefont {J.}~\bibnamefont {Aarts}},\ }\href {\doibase
  10.1103/PhysRevLett.100.047002} {\bibfield  {journal} {\bibinfo  {journal}
  {Phys. Rev. Lett.}\ }\textbf {\bibinfo {volume} {100}},\ \bibinfo {pages}
  {047002} (\bibinfo {year} {2008})}\BibitemShut {NoStop}%
\bibitem [{\citenamefont {Jeon}\ \emph
  {et~al.}(2019{\natexlab{a}})\citenamefont {Jeon}, \citenamefont {Ciccarelli},
  \citenamefont {Kurebayashi}, \citenamefont {Cohen}, \citenamefont {Komori},
  \citenamefont {Robinson},\ and\ \citenamefont {Blamire}}]{Jeon2019}%
  \BibitemOpen
  \bibfield  {author} {\bibinfo {author} {\bibfnamefont {K.-R.}\ \bibnamefont
  {Jeon}}, \bibinfo {author} {\bibfnamefont {C.}~\bibnamefont {Ciccarelli}},
  \bibinfo {author} {\bibfnamefont {H.}~\bibnamefont {Kurebayashi}}, \bibinfo
  {author} {\bibfnamefont {L.~F.}\ \bibnamefont {Cohen}}, \bibinfo {author}
  {\bibfnamefont {S.}~\bibnamefont {Komori}}, \bibinfo {author} {\bibfnamefont
  {J.~W.~A.}\ \bibnamefont {Robinson}}, \ and\ \bibinfo {author} {\bibfnamefont
  {M.~G.}\ \bibnamefont {Blamire}},\ }\href {\doibase
  10.1103/PhysRevB.99.144503} {\bibfield  {journal} {\bibinfo  {journal} {Phys.
  Rev. B}\ }\textbf {\bibinfo {volume} {99}},\ \bibinfo {pages} {144503}
  (\bibinfo {year} {2019}{\natexlab{a}})}\BibitemShut {NoStop}%
\bibitem [{\citenamefont {Jeon}\ \emph
  {et~al.}(2019{\natexlab{b}})\citenamefont {Jeon}, \citenamefont {Ciccarelli},
  \citenamefont {Kurebayashi}, \citenamefont {Cohen}, \citenamefont {Montiel},
  \citenamefont {Eschrig}, \citenamefont {Komori}, \citenamefont {Robinson},\
  and\ \citenamefont {Blamire}}]{Jeon2019_2}%
  \BibitemOpen
  \bibfield  {author} {\bibinfo {author} {\bibfnamefont {K.-R.}\ \bibnamefont
  {Jeon}}, \bibinfo {author} {\bibfnamefont {C.}~\bibnamefont {Ciccarelli}},
  \bibinfo {author} {\bibfnamefont {H.}~\bibnamefont {Kurebayashi}}, \bibinfo
  {author} {\bibfnamefont {L.~F.}\ \bibnamefont {Cohen}}, \bibinfo {author}
  {\bibfnamefont {X.}~\bibnamefont {Montiel}}, \bibinfo {author} {\bibfnamefont
  {M.}~\bibnamefont {Eschrig}}, \bibinfo {author} {\bibfnamefont
  {S.}~\bibnamefont {Komori}}, \bibinfo {author} {\bibfnamefont {J.~W.~A.}\
  \bibnamefont {Robinson}}, \ and\ \bibinfo {author} {\bibfnamefont {M.~G.}\
  \bibnamefont {Blamire}},\ }\href {\doibase 10.1103/PhysRevB.99.024507}
  {\bibfield  {journal} {\bibinfo  {journal} {Phys. Rev. B}\ }\textbf {\bibinfo
  {volume} {99}},\ \bibinfo {pages} {024507} (\bibinfo {year}
  {2019}{\natexlab{b}})}\BibitemShut {NoStop}%
\bibitem [{\citenamefont {Jeon}\ \emph
  {et~al.}(2019{\natexlab{c}})\citenamefont {Jeon}, \citenamefont {Ciccarelli},
  \citenamefont {Kurebayashi}, \citenamefont {Cohen}, \citenamefont {Montiel},
  \citenamefont {Eschrig}, \citenamefont {Wagner}, \citenamefont {Komori},
  \citenamefont {Srivastava}, \citenamefont {Robinson},\ and\ \citenamefont
  {Blamire}}]{Jeon2019_3}%
  \BibitemOpen
  \bibfield  {author} {\bibinfo {author} {\bibfnamefont {K.-R.}\ \bibnamefont
  {Jeon}}, \bibinfo {author} {\bibfnamefont {C.}~\bibnamefont {Ciccarelli}},
  \bibinfo {author} {\bibfnamefont {H.}~\bibnamefont {Kurebayashi}}, \bibinfo
  {author} {\bibfnamefont {L.~F.}\ \bibnamefont {Cohen}}, \bibinfo {author}
  {\bibfnamefont {X.}~\bibnamefont {Montiel}}, \bibinfo {author} {\bibfnamefont
  {M.}~\bibnamefont {Eschrig}}, \bibinfo {author} {\bibfnamefont
  {T.}~\bibnamefont {Wagner}}, \bibinfo {author} {\bibfnamefont
  {S.}~\bibnamefont {Komori}}, \bibinfo {author} {\bibfnamefont
  {A.}~\bibnamefont {Srivastava}}, \bibinfo {author} {\bibfnamefont {J.~W.}\
  \bibnamefont {Robinson}}, \ and\ \bibinfo {author} {\bibfnamefont {M.~G.}\
  \bibnamefont {Blamire}},\ }\href {\doibase 10.1103/PhysRevApplied.11.014061}
  {\bibfield  {journal} {\bibinfo  {journal} {Phys. Rev. Applied}\ }\textbf
  {\bibinfo {volume} {11}},\ \bibinfo {pages} {014061} (\bibinfo {year}
  {2019}{\natexlab{c}})}\BibitemShut {NoStop}%
\bibitem [{\citenamefont {Jeon}\ \emph {et~al.}(2018)\citenamefont {Jeon},
  \citenamefont {Ciccarelli}, \citenamefont {Ferguson}, \citenamefont
  {Kurebayashi}, \citenamefont {Cohen}, \citenamefont {Montiel}, \citenamefont
  {Eschrig}, \citenamefont {Robinson},\ and\ \citenamefont
  {Blamire}}]{Jeon2018}%
  \BibitemOpen
  \bibfield  {author} {\bibinfo {author} {\bibfnamefont {K.-R.}\ \bibnamefont
  {Jeon}}, \bibinfo {author} {\bibfnamefont {C.}~\bibnamefont {Ciccarelli}},
  \bibinfo {author} {\bibfnamefont {A.~J.}\ \bibnamefont {Ferguson}}, \bibinfo
  {author} {\bibfnamefont {H.}~\bibnamefont {Kurebayashi}}, \bibinfo {author}
  {\bibfnamefont {L.~F.}\ \bibnamefont {Cohen}}, \bibinfo {author}
  {\bibfnamefont {X.}~\bibnamefont {Montiel}}, \bibinfo {author} {\bibfnamefont
  {M.}~\bibnamefont {Eschrig}}, \bibinfo {author} {\bibfnamefont {J.~W.~A.}\
  \bibnamefont {Robinson}}, \ and\ \bibinfo {author} {\bibfnamefont {M.~G.}\
  \bibnamefont {Blamire}},\ }\href {\doibase 10.1038/s41563-018-0058-9}
  {\bibfield  {journal} {\bibinfo  {journal} {Nature Materials}\ }\textbf
  {\bibinfo {volume} {17}},\ \bibinfo {pages} {499} (\bibinfo {year}
  {2018})}\BibitemShut {NoStop}%
\bibitem [{\citenamefont {Yao}\ \emph {et~al.}(2018)\citenamefont {Yao},
  \citenamefont {Song}, \citenamefont {Takamura}, \citenamefont {Cascales},
  \citenamefont {Yuan}, \citenamefont {Ma}, \citenamefont {Yun}, \citenamefont
  {Xie}, \citenamefont {Moodera},\ and\ \citenamefont {Han}}]{Yao2018}%
  \BibitemOpen
  \bibfield  {author} {\bibinfo {author} {\bibfnamefont {Y.}~\bibnamefont
  {Yao}}, \bibinfo {author} {\bibfnamefont {Q.}~\bibnamefont {Song}}, \bibinfo
  {author} {\bibfnamefont {Y.}~\bibnamefont {Takamura}}, \bibinfo {author}
  {\bibfnamefont {J.~P.}\ \bibnamefont {Cascales}}, \bibinfo {author}
  {\bibfnamefont {W.}~\bibnamefont {Yuan}}, \bibinfo {author} {\bibfnamefont
  {Y.}~\bibnamefont {Ma}}, \bibinfo {author} {\bibfnamefont {Y.}~\bibnamefont
  {Yun}}, \bibinfo {author} {\bibfnamefont {X.~C.}\ \bibnamefont {Xie}},
  \bibinfo {author} {\bibfnamefont {J.~S.}\ \bibnamefont {Moodera}}, \ and\
  \bibinfo {author} {\bibfnamefont {W.}~\bibnamefont {Han}},\ }\href {\doibase
  10.1103/PhysRevB.97.224414} {\bibfield  {journal} {\bibinfo  {journal} {Phys.
  Rev. B}\ }\textbf {\bibinfo {volume} {97}},\ \bibinfo {pages} {224414}
  (\bibinfo {year} {2018})}\BibitemShut {NoStop}%
\bibitem [{\citenamefont {Li}\ \emph {et~al.}(2018)\citenamefont {Li},
  \citenamefont {Zhao}, \citenamefont {Zhang},\ and\ \citenamefont
  {Sun}}]{Li2018}%
  \BibitemOpen
  \bibfield  {author} {\bibinfo {author} {\bibfnamefont {L.~L.}\ \bibnamefont
  {Li}}, \bibinfo {author} {\bibfnamefont {Y.~L.}\ \bibnamefont {Zhao}},
  \bibinfo {author} {\bibfnamefont {X.~X.}\ \bibnamefont {Zhang}}, \ and\
  \bibinfo {author} {\bibfnamefont {Y.}~\bibnamefont {Sun}},\ }\href@noop {}
  {\bibfield  {journal} {\bibinfo  {journal} {Chin. Phys. Lett.}\ }\textbf
  {\bibinfo {volume} {35}},\ \bibinfo {pages} {077401} (\bibinfo {year}
  {2018})}\BibitemShut {NoStop}%
\bibitem [{\citenamefont {Golovchanskiy}\ \emph {et~al.}(2020)\citenamefont
  {Golovchanskiy}, \citenamefont {Abramov}, \citenamefont {Stolyarov},
  \citenamefont {Chichkov}, \citenamefont {Silaev}, \citenamefont {Shchetinin},
  \citenamefont {Golubov}, \citenamefont {Ryazanov}, \citenamefont {Ustinov},\
  and\ \citenamefont {Kupriyanov}}]{Golovchanskiy2020}%
  \BibitemOpen
  \bibfield  {author} {\bibinfo {author} {\bibfnamefont {I.}~\bibnamefont
  {Golovchanskiy}}, \bibinfo {author} {\bibfnamefont {N.}~\bibnamefont
  {Abramov}}, \bibinfo {author} {\bibfnamefont {V.}~\bibnamefont {Stolyarov}},
  \bibinfo {author} {\bibfnamefont {V.}~\bibnamefont {Chichkov}}, \bibinfo
  {author} {\bibfnamefont {M.}~\bibnamefont {Silaev}}, \bibinfo {author}
  {\bibfnamefont {I.}~\bibnamefont {Shchetinin}}, \bibinfo {author}
  {\bibfnamefont {A.}~\bibnamefont {Golubov}}, \bibinfo {author} {\bibfnamefont
  {V.}~\bibnamefont {Ryazanov}}, \bibinfo {author} {\bibfnamefont
  {A.}~\bibnamefont {Ustinov}}, \ and\ \bibinfo {author} {\bibfnamefont
  {M.}~\bibnamefont {Kupriyanov}},\ }\href {\doibase
  10.1103/PhysRevApplied.14.024086} {\bibfield  {journal} {\bibinfo  {journal}
  {Phys. Rev. Applied}\ }\textbf {\bibinfo {volume} {14}},\ \bibinfo {pages}
  {024086} (\bibinfo {year} {2020})}\BibitemShut {NoStop}%
\bibitem [{\citenamefont {Silaev}(2020)}]{Silaev2020}%
  \BibitemOpen
  \bibfield  {author} {\bibinfo {author} {\bibfnamefont {M.~A.}\ \bibnamefont
  {Silaev}},\ }\href {\doibase 10.1103/PhysRevB.102.144521} {\bibfield
  {journal} {\bibinfo  {journal} {Phys. Rev. B}\ }\textbf {\bibinfo {volume}
  {102}},\ \bibinfo {pages} {144521} (\bibinfo {year} {2020})}\BibitemShut
  {NoStop}%
\bibitem [{\citenamefont {Dobrovolskiy}\ \emph {et~al.}(2019)\citenamefont
  {Dobrovolskiy}, \citenamefont {Sachser}, \citenamefont {Br{\"a}cher},
  \citenamefont {B{\"o}ttcher}, \citenamefont {Kruglyak}, \citenamefont {Vovk},
  \citenamefont {Shklovskij}, \citenamefont {Huth}, \citenamefont
  {Hillebrands},\ and\ \citenamefont {Chumak}}]{Dobrovolskiy2019}%
  \BibitemOpen
  \bibfield  {author} {\bibinfo {author} {\bibfnamefont {O.~V.}\ \bibnamefont
  {Dobrovolskiy}}, \bibinfo {author} {\bibfnamefont {R.}~\bibnamefont
  {Sachser}}, \bibinfo {author} {\bibfnamefont {T.}~\bibnamefont
  {Br{\"a}cher}}, \bibinfo {author} {\bibfnamefont {T.}~\bibnamefont
  {B{\"o}ttcher}}, \bibinfo {author} {\bibfnamefont {V.~V.}\ \bibnamefont
  {Kruglyak}}, \bibinfo {author} {\bibfnamefont {R.~V.}\ \bibnamefont {Vovk}},
  \bibinfo {author} {\bibfnamefont {V.~A.}\ \bibnamefont {Shklovskij}},
  \bibinfo {author} {\bibfnamefont {M.}~\bibnamefont {Huth}}, \bibinfo {author}
  {\bibfnamefont {B.}~\bibnamefont {Hillebrands}}, \ and\ \bibinfo {author}
  {\bibfnamefont {A.~V.}\ \bibnamefont {Chumak}},\ }\href {\doibase
  10.1038/s41567-019-0428-5} {\bibfield  {journal} {\bibinfo  {journal} {Nature
  Physics}\ }\textbf {\bibinfo {volume} {15}},\ \bibinfo {pages} {477}
  (\bibinfo {year} {2019})}\BibitemShut {NoStop}%
\bibitem [{\citenamefont {Yu}\ and\ \citenamefont {Bauer}(2022)}]{Yu2022}%
  \BibitemOpen
  \bibfield  {author} {\bibinfo {author} {\bibfnamefont {T.}~\bibnamefont
  {Yu}}\ and\ \bibinfo {author} {\bibfnamefont {G.~E.~W.}\ \bibnamefont
  {Bauer}},\ }\href {\doibase 10.1103/PhysRevLett.129.117201} {\bibfield
  {journal} {\bibinfo  {journal} {Phys. Rev. Lett.}\ }\textbf {\bibinfo
  {volume} {129}},\ \bibinfo {pages} {117201} (\bibinfo {year}
  {2022})}\BibitemShut {NoStop}%
\bibitem [{\citenamefont {Bobkova}\ \emph {et~al.}(2022)\citenamefont
  {Bobkova}, \citenamefont {Bobkov}, \citenamefont {Kamra},\ and\ \citenamefont
  {Belzig}}]{Bobkova2022}%
  \BibitemOpen
  \bibfield  {author} {\bibinfo {author} {\bibfnamefont {I.~V.}\ \bibnamefont
  {Bobkova}}, \bibinfo {author} {\bibfnamefont {A.~M.}\ \bibnamefont {Bobkov}},
  \bibinfo {author} {\bibfnamefont {A.}~\bibnamefont {Kamra}}, \ and\ \bibinfo
  {author} {\bibfnamefont {W.}~\bibnamefont {Belzig}},\ }\href {\doibase
  10.1038/s43246-022-00321-8} {\bibfield  {journal} {\bibinfo  {journal}
  {Communications Materials}\ }\textbf {\bibinfo {volume} {3}},\ \bibinfo
  {pages} {95} (\bibinfo {year} {2022})}\BibitemShut {NoStop}%
\bibitem [{\citenamefont {Rohling}\ \emph {et~al.}(2018)\citenamefont
  {Rohling}, \citenamefont {Fj\ae{}rbu},\ and\ \citenamefont
  {Brataas}}]{Rohling2018}%
  \BibitemOpen
  \bibfield  {author} {\bibinfo {author} {\bibfnamefont {N.}~\bibnamefont
  {Rohling}}, \bibinfo {author} {\bibfnamefont {E.~L.}\ \bibnamefont
  {Fj\ae{}rbu}}, \ and\ \bibinfo {author} {\bibfnamefont {A.}~\bibnamefont
  {Brataas}},\ }\href {\doibase 10.1103/PhysRevB.97.115401} {\bibfield
  {journal} {\bibinfo  {journal} {Phys. Rev. B}\ }\textbf {\bibinfo {volume}
  {97}},\ \bibinfo {pages} {115401} (\bibinfo {year} {2018})}\BibitemShut
  {NoStop}%
\bibitem [{\citenamefont {Fj\ae{}rbu}\ \emph {et~al.}(2019)\citenamefont
  {Fj\ae{}rbu}, \citenamefont {Rohling},\ and\ \citenamefont
  {Brataas}}]{Fjaerbu2019}%
  \BibitemOpen
  \bibfield  {author} {\bibinfo {author} {\bibfnamefont {E.~L.}\ \bibnamefont
  {Fj\ae{}rbu}}, \bibinfo {author} {\bibfnamefont {N.}~\bibnamefont {Rohling}},
  \ and\ \bibinfo {author} {\bibfnamefont {A.}~\bibnamefont {Brataas}},\ }\href
  {\doibase 10.1103/PhysRevB.100.125432} {\bibfield  {journal} {\bibinfo
  {journal} {Phys. Rev. B}\ }\textbf {\bibinfo {volume} {100}},\ \bibinfo
  {pages} {125432} (\bibinfo {year} {2019})}\BibitemShut {NoStop}%
\bibitem [{\citenamefont {Erlandsen}\ \emph {et~al.}(2019)\citenamefont
  {Erlandsen}, \citenamefont {Kamra}, \citenamefont {Brataas},\ and\
  \citenamefont {Sudb\o{}}}]{Erlandsen2019}%
  \BibitemOpen
  \bibfield  {author} {\bibinfo {author} {\bibfnamefont {E.}~\bibnamefont
  {Erlandsen}}, \bibinfo {author} {\bibfnamefont {A.}~\bibnamefont {Kamra}},
  \bibinfo {author} {\bibfnamefont {A.}~\bibnamefont {Brataas}}, \ and\
  \bibinfo {author} {\bibfnamefont {A.}~\bibnamefont {Sudb\o{}}},\ }\href
  {\doibase 10.1103/PhysRevB.100.100503} {\bibfield  {journal} {\bibinfo
  {journal} {Phys. Rev. B}\ }\textbf {\bibinfo {volume} {100}},\ \bibinfo
  {pages} {100503} (\bibinfo {year} {2019})}\BibitemShut {NoStop}%
\bibitem [{\citenamefont {Thingstad}\ \emph {et~al.}(2021)\citenamefont
  {Thingstad}, \citenamefont {Erlandsen},\ and\ \citenamefont
  {Sudb\o{}}}]{Thingstad2021}%
  \BibitemOpen
  \bibfield  {author} {\bibinfo {author} {\bibfnamefont {E.}~\bibnamefont
  {Thingstad}}, \bibinfo {author} {\bibfnamefont {E.}~\bibnamefont
  {Erlandsen}}, \ and\ \bibinfo {author} {\bibfnamefont {A.}~\bibnamefont
  {Sudb\o{}}},\ }\href {\doibase 10.1103/PhysRevB.104.014508} {\bibfield
  {journal} {\bibinfo  {journal} {Phys. Rev. B}\ }\textbf {\bibinfo {volume}
  {104}},\ \bibinfo {pages} {014508} (\bibinfo {year} {2021})}\BibitemShut
  {NoStop}%
\bibitem [{\citenamefont {M\ae{}land}\ and\ \citenamefont
  {Sudb\o{}}(2023)}]{Maeland2023}%
  \BibitemOpen
  \bibfield  {author} {\bibinfo {author} {\bibfnamefont {K.}~\bibnamefont
  {M\ae{}land}}\ and\ \bibinfo {author} {\bibfnamefont {A.}~\bibnamefont
  {Sudb\o{}}},\ }\href {\doibase 10.1103/PhysRevLett.130.156002} {\bibfield
  {journal} {\bibinfo  {journal} {Phys. Rev. Lett.}\ }\textbf {\bibinfo
  {volume} {130}},\ \bibinfo {pages} {156002} (\bibinfo {year}
  {2023})}\BibitemShut {NoStop}%
\bibitem [{\citenamefont {Gong}\ \emph {et~al.}(XXXX)\citenamefont {Gong},
  \citenamefont {Kargarian}, \citenamefont {Stern}, \citenamefont {Yue},
  \citenamefont {Zhou}, \citenamefont {Jin}, \citenamefont {Galitski},
  \citenamefont {Yakovenko},\ and\ \citenamefont {Xia}}]{Gong2023}%
  \BibitemOpen
  \bibfield  {author} {\bibinfo {author} {\bibfnamefont {X.}~\bibnamefont
  {Gong}}, \bibinfo {author} {\bibfnamefont {M.}~\bibnamefont {Kargarian}},
  \bibinfo {author} {\bibfnamefont {A.}~\bibnamefont {Stern}}, \bibinfo
  {author} {\bibfnamefont {D.}~\bibnamefont {Yue}}, \bibinfo {author}
  {\bibfnamefont {H.}~\bibnamefont {Zhou}}, \bibinfo {author} {\bibfnamefont
  {X.}~\bibnamefont {Jin}}, \bibinfo {author} {\bibfnamefont {V.~M.}\
  \bibnamefont {Galitski}}, \bibinfo {author} {\bibfnamefont {V.~M.}\
  \bibnamefont {Yakovenko}}, \ and\ \bibinfo {author} {\bibfnamefont
  {J.}~\bibnamefont {Xia}},\ }\href {\doibase 10.1126/sciadv.1602579}
  {\bibfield  {journal} {\bibinfo  {journal} {Science Advances}\ }\textbf
  {\bibinfo {volume} {3}},\ \bibinfo {pages} {e1602579} (\bibinfo {year}
  {XXXX})}\BibitemShut {NoStop}%
\bibitem [{\citenamefont {Auslender}\ and\ \citenamefont
  {Irkhin}(1985)}]{Auslender1985}%
  \BibitemOpen
  \bibfield  {author} {\bibinfo {author} {\bibfnamefont {M.~I.}\ \bibnamefont
  {Auslender}}\ and\ \bibinfo {author} {\bibfnamefont {V.}~\bibnamefont
  {Irkhin}},\ }\href
  {https://www.sciencedirect.com/science/article/pii/0038109885907823}
  {\bibfield  {journal} {\bibinfo  {journal} {Solid State Communications}\
  }\textbf {\bibinfo {volume} {56}},\ \bibinfo {pages} {701} (\bibinfo {year}
  {1985})}\BibitemShut {NoStop}%
\bibitem [{\citenamefont {Appelbaum}\ and\ \citenamefont
  {Brinkman}(1969)}]{Appelbaum1969}%
  \BibitemOpen
  \bibfield  {author} {\bibinfo {author} {\bibfnamefont {J.~A.}\ \bibnamefont
  {Appelbaum}}\ and\ \bibinfo {author} {\bibfnamefont {W.~F.}\ \bibnamefont
  {Brinkman}},\ }\href {\doibase 10.1103/PhysRev.183.553} {\bibfield  {journal}
  {\bibinfo  {journal} {Phys. Rev.}\ }\textbf {\bibinfo {volume} {183}},\
  \bibinfo {pages} {553} (\bibinfo {year} {1969})}\BibitemShut {NoStop}%
\bibitem [{\citenamefont {Kittel}(1963)}]{Kittel}%
  \BibitemOpen
  \bibfield  {author} {\bibinfo {author} {\bibfnamefont {C.}~\bibnamefont
  {Kittel}},\ }\href@noop {} {\emph {\bibinfo {title} {Quantum theory of
  solids}}}\ (\bibinfo  {publisher} {Wiley},\ \bibinfo {address} {New York},\
  \bibinfo {year} {1963})\BibitemShut {NoStop}%
\bibitem [{\citenamefont {Holstein}\ and\ \citenamefont
  {Primakoff}(1940)}]{HP1940}%
  \BibitemOpen
  \bibfield  {author} {\bibinfo {author} {\bibfnamefont {T.}~\bibnamefont
  {Holstein}}\ and\ \bibinfo {author} {\bibfnamefont {H.}~\bibnamefont
  {Primakoff}},\ }\href {\doibase 10.1103/PhysRev.58.1098} {\bibfield
  {journal} {\bibinfo  {journal} {Phys. Rev.}\ }\textbf {\bibinfo {volume}
  {58}},\ \bibinfo {pages} {1098} (\bibinfo {year} {1940})}\BibitemShut
  {NoStop}%
\bibitem [{\citenamefont {Kamra}\ and\ \citenamefont
  {Belzig}(2016)}]{Kamra2016}%
  \BibitemOpen
  \bibfield  {author} {\bibinfo {author} {\bibfnamefont {A.}~\bibnamefont
  {Kamra}}\ and\ \bibinfo {author} {\bibfnamefont {W.}~\bibnamefont {Belzig}},\
  }\href {\doibase 10.1103/PhysRevB.94.014419} {\bibfield  {journal} {\bibinfo
  {journal} {Phys. Rev. B}\ }\textbf {\bibinfo {volume} {94}},\ \bibinfo
  {pages} {014419} (\bibinfo {year} {2016})}\BibitemShut {NoStop}%
\bibitem [{\citenamefont {Hijano}\ \emph {et~al.}(2021)\citenamefont {Hijano},
  \citenamefont {Ili\ifmmode~\acute{c}\else \'{c}\fi{}}, \citenamefont {Rouco},
  \citenamefont {Gonz\'alez-Orellana}, \citenamefont {Ilyn}, \citenamefont
  {Rogero}, \citenamefont {Virtanen}, \citenamefont {Heikkil\"a}, \citenamefont
  {Khorshidian}, \citenamefont {Spies}, \citenamefont {Ligato}, \citenamefont
  {Giazotto}, \citenamefont {Strambini},\ and\ \citenamefont
  {Bergeret}}]{Hijano2021}%
  \BibitemOpen
  \bibfield  {author} {\bibinfo {author} {\bibfnamefont {A.}~\bibnamefont
  {Hijano}}, \bibinfo {author} {\bibfnamefont {S.}~\bibnamefont
  {Ili\ifmmode~\acute{c}\else \'{c}\fi{}}}, \bibinfo {author} {\bibfnamefont
  {M.}~\bibnamefont {Rouco}}, \bibinfo {author} {\bibfnamefont
  {C.}~\bibnamefont {Gonz\'alez-Orellana}}, \bibinfo {author} {\bibfnamefont
  {M.}~\bibnamefont {Ilyn}}, \bibinfo {author} {\bibfnamefont {C.}~\bibnamefont
  {Rogero}}, \bibinfo {author} {\bibfnamefont {P.}~\bibnamefont {Virtanen}},
  \bibinfo {author} {\bibfnamefont {T.~T.}\ \bibnamefont {Heikkil\"a}},
  \bibinfo {author} {\bibfnamefont {S.}~\bibnamefont {Khorshidian}}, \bibinfo
  {author} {\bibfnamefont {M.}~\bibnamefont {Spies}}, \bibinfo {author}
  {\bibfnamefont {N.}~\bibnamefont {Ligato}}, \bibinfo {author} {\bibfnamefont
  {F.}~\bibnamefont {Giazotto}}, \bibinfo {author} {\bibfnamefont
  {E.}~\bibnamefont {Strambini}}, \ and\ \bibinfo {author} {\bibfnamefont
  {F.~S.}\ \bibnamefont {Bergeret}},\ }\href {\doibase
  10.1103/PhysRevResearch.3.023131} {\bibfield  {journal} {\bibinfo  {journal}
  {Phys. Rev. Res.}\ }\textbf {\bibinfo {volume} {3}},\ \bibinfo {pages}
  {023131} (\bibinfo {year} {2021})}\BibitemShut {NoStop}%
\bibitem [{\citenamefont {Kamra}\ \emph {et~al.}(2018)\citenamefont {Kamra},
  \citenamefont {Rezaei},\ and\ \citenamefont {Belzig}}]{Kamra2018}%
  \BibitemOpen
  \bibfield  {author} {\bibinfo {author} {\bibfnamefont {A.}~\bibnamefont
  {Kamra}}, \bibinfo {author} {\bibfnamefont {A.}~\bibnamefont {Rezaei}}, \
  and\ \bibinfo {author} {\bibfnamefont {W.}~\bibnamefont {Belzig}},\ }\href
  {\doibase 10.1103/PhysRevLett.121.247702} {\bibfield  {journal} {\bibinfo
  {journal} {Phys. Rev. Lett.}\ }\textbf {\bibinfo {volume} {121}},\ \bibinfo
  {pages} {247702} (\bibinfo {year} {2018})}\BibitemShut {NoStop}%
\bibitem [{\citenamefont {Xiao}\ \emph {et~al.}(2010)\citenamefont {Xiao},
  \citenamefont {Bauer}, \citenamefont {Uchida}, \citenamefont {Saitoh},\ and\
  \citenamefont {Maekawa}}]{Xiao2010}%
  \BibitemOpen
  \bibfield  {author} {\bibinfo {author} {\bibfnamefont {J.}~\bibnamefont
  {Xiao}}, \bibinfo {author} {\bibfnamefont {G.~E.~W.}\ \bibnamefont {Bauer}},
  \bibinfo {author} {\bibfnamefont {K.-c.}\ \bibnamefont {Uchida}}, \bibinfo
  {author} {\bibfnamefont {E.}~\bibnamefont {Saitoh}}, \ and\ \bibinfo {author}
  {\bibfnamefont {S.}~\bibnamefont {Maekawa}},\ }\href {\doibase
  10.1103/PhysRevB.81.214418} {\bibfield  {journal} {\bibinfo  {journal} {Phys.
  Rev. B}\ }\textbf {\bibinfo {volume} {81}},\ \bibinfo {pages} {214418}
  (\bibinfo {year} {2010})}\BibitemShut {NoStop}%
\bibitem [{\citenamefont {Wakamura}\ \emph {et~al.}(2014)\citenamefont
  {Wakamura}, \citenamefont {Hasegawa}, \citenamefont {Ohnishi}, \citenamefont
  {Niimi},\ and\ \citenamefont {Otani}}]{Wakamura2014}%
  \BibitemOpen
  \bibfield  {author} {\bibinfo {author} {\bibfnamefont {T.}~\bibnamefont
  {Wakamura}}, \bibinfo {author} {\bibfnamefont {N.}~\bibnamefont {Hasegawa}},
  \bibinfo {author} {\bibfnamefont {K.}~\bibnamefont {Ohnishi}}, \bibinfo
  {author} {\bibfnamefont {Y.}~\bibnamefont {Niimi}}, \ and\ \bibinfo {author}
  {\bibfnamefont {Y.}~\bibnamefont {Otani}},\ }\href {\doibase
  10.1103/PhysRevLett.112.036602} {\bibfield  {journal} {\bibinfo  {journal}
  {Phys. Rev. Lett.}\ }\textbf {\bibinfo {volume} {112}},\ \bibinfo {pages}
  {036602} (\bibinfo {year} {2014})}\BibitemShut {NoStop}%
\bibitem [{\citenamefont {B\"oni}\ \emph {et~al.}(1995)\citenamefont {B\"oni},
  \citenamefont {Hennion},\ and\ \citenamefont {Mart\'{\i}nez}}]{Boni1995}%
  \BibitemOpen
  \bibfield  {author} {\bibinfo {author} {\bibfnamefont {P.}~\bibnamefont
  {B\"oni}}, \bibinfo {author} {\bibfnamefont {M.}~\bibnamefont {Hennion}}, \
  and\ \bibinfo {author} {\bibfnamefont {J.~L.}\ \bibnamefont
  {Mart\'{\i}nez}},\ }\href {\doibase 10.1103/PhysRevB.52.10142} {\bibfield
  {journal} {\bibinfo  {journal} {Phys. Rev. B}\ }\textbf {\bibinfo {volume}
  {52}},\ \bibinfo {pages} {10142} (\bibinfo {year} {1995})}\BibitemShut
  {NoStop}%
\bibitem [{\citenamefont {Getzlaff}(2008)}]{Getzlaff}%
  \BibitemOpen
  \bibfield  {author} {\bibinfo {author} {\bibfnamefont {M.}~\bibnamefont
  {Getzlaff}},\ }\href@noop {} {\emph {\bibinfo {title} {Fundamentals of
  magnetism}}}\ (\bibinfo  {publisher} {Springer-Verlag},\ \bibinfo {address}
  {Berlin Heidelberg},\ \bibinfo {year} {2008})\BibitemShut {NoStop}%
\end{thebibliography}%

\end{document}